\documentclass[12pt]{iopart}

\usepackage{graphicx}  
\usepackage{subfig}
\begin{document}

\title[]{Study of nuclear matter properties for Hybrid EoS}

\author{Ishfaq A. Rather$^1$, A. A. Usmani$^1$, S. K. Patra$^{2,3}$}

\address{$^1$Department of Physics, Aligarh Muslim University, Aligarh 202002, India}
\address{$^2$Institute of Physics, Bhubaneswar 751005, India}
\address{$^3$ Homi Bhabha National Institute, Training School Complex, Anushakti Nagar, Mumbai 400085, India}
\ead{ishfaqrather81@gmail.com}
\vspace{10pt}
\begin{indented}
\item \noindent keywords: {\it  Symmetry energy, Equation of State, hybrid eos\/}
\end{indented}

\begin{abstract}
	We study the nuclear matter properties like symmetry energy, slope parameter, curvature, skewness and incompressibility for Hybrid EoS. The hybrid EoS is constructed by combining the hadron phase with the quark phase. For the hadron phase, we use the recently proposed Effective-Field-Theory motivated Relativistic Mean-Field  model (E-RMF) with different parameter sets. For the quark phase, we employ the simple MIT Bag model with different Bag constants. The mixed phase formed by the hadron-quark phase transition is studied using the Gibbs construction. The nuclear matter properties for hybrid EoS are calculated and their variation with the bag constant is determined. Star matter properties like mass and radius are also calculated for the obtained EoS.
\end{abstract}

%
%
%
%
%

\section{Introduction}

Since the Quark matter is by assumption completely stable, it may be the true ground state of the hadronic matter \cite{PhysRevD.30.272,PhysRevD.30.2379}. So the quark matter, the deconfined quark phase, is quite likely to occur in the inner regions of the  compact objects like neutron stars. It may exist both as a pure phase in the central regions and as a mixed phase with hadronic matter \cite{PhysRevD.46.1274}. The neutron stars with a hadronic crust and a quark core (pure or mixed) are termed as Hybrid stars.\par 
In nuclear physics and nuclear astrophysics, the Equation of State (EoS) plays a very crucial rule in understanding the nature of matter in finite and infinite nuclear matter \cite{Danielewicz1592,Lattimer:2015nhk,RevModPhys.88.021001,RevModPhys.89.015007,Ozel:2016oaf}. The binding energy per nucleon $e(\rho,\alpha$) = $\mathcal{E}/A$ and the isospin asymmetry $\alpha$= $(\rho_n - \rho_p)/\rho$ are one of the basic inputs for calculating the pressure and the energy density (EoS) of neutron star matter. The symmetry energy $S(\rho)$ and other quantities have a huge impact on the EoS. However, the $S(\rho)$ cannot be measured directly, so fully depends on the theoretical models. Unfortunately, these models predict a wide range of symmetry energy \cite{PhysRevC.85.035201,PhysRevC.90.055203}. At saturation density $(\rho_0)$, all these quantities are known more or less to a good extent, but the results are very much uncertain for the densities above $\rho_0$. While many theoretical models predict the symmetry energy $S(\rho)$ to be increasing with the density, several other models predict that the $S(\rho)$ increases with the density upto $\rho_0$ and thereafter decreases \cite{PhysRevC.90.055203,PhysRevLett.106.252501,PhysRevC.85.035201}.  At densities around 2-8 $\rho_0$, the symmetry energy and the higher derivatives such as the slope parameter $L=3\rho_0 S'(\rho_0)$, curvature of symmetry energy $K_{sym}=9\rho_0^2 S''(\rho_0)$, $Q_{sym}=27\rho_0^3 S'''(\rho_0)$ and also the incompressibility plays a key role in determining the structure and properties of neutron stars \cite{PhysRevC.98.065801} and the possibility of the exotic phases \cite{THORSSON1994693,PhysRevC.95.015801}. \par 
The properties of neutron star such as its composition, mass, radius etc. depend upon the equation of state. The outer part of the neutron star where the density is low ($\approx \rho_0$) is mainly described by the hadronic matter. As the density increases (5-10 $\rho_0$), a phase transition from hadronic matter to quark matter is possible, where a mixed hadron-quark phase is formed for a certain density range followed a pure quark phase.\par 
In the present work, we combine the two phases together to build a single hybrid EoS. We calculate the nuclear matter properties for hybrid EoS and the effect of bag constant on these properties. \par 
This paper is organized as follows: in section 2 we discuss the theoretical approaches employed to study the equation of state of different phases. For hadronic matter, the Effective-field -theory motivated Relativistic Mean-Field (E-RMF) model\cite{FURNSTAHL1996539} is employed by using recently proposed different parameter sets. In the Quark matter, the MIT Bag Model is used to describe the Unpaired Quark Matter (UQM) \cite{PhysRevD.9.3471,PhysRevD.17.1109,PhysRevD.30.2379}. The mixed phase for the hybrid EoS is obtained by using the Gibbs construction \cite{PhysRevD.46.1274}, where the mixed phase follows global charge neutrality conditions. In section 3, we discuss the nuclear matter (NM) properties like symmetry energy and other quantities that play crucial role in studying the EoS. We then calculate the NM properties for this hybrid EoS. All the calculated results are discussed in section 4.  The mass-radius profiles for the obtained EoS's are also shown Finally the summary and conclusions are given in section 5.

\section{Formalism}
\label{sec:headings}

\subsection*{{\textbf A. Hadron Matter}}
Quantum Hadrodynamics (QHD), the Effective Field Theory (EFT) for strong interaction \cite{Walecka:1974qa,Reinhard:1989zi,Serot_1992} at low energies has been studied extensively to describe the properties of both finite nuclei \cite{Horowitz:1981xw,BOGUTA1977413,GAMBHIR1990132,RING1996193} and infinite nuclear matter \cite{Walecka:1974qa,Arumugam:2004ys}.In this theory, the interaction of nucleons occurs with the exchange of mesons like $\sigma$, $\omega$, $\rho$ and $\delta$.\par 
The basic relativistic Lagrangian has the contribution from $\sigma, \omega$ and $\rho$ mesons without any self-coupling terms which is the original Walecka model \cite{Horowitz:1981xw}. The prediction of the nuclear incompressibility $K$ by this model is very large ($\approx$ 550 MeV) \cite{Walecka:1974qa} and hence the self-coupling terms were added by Boguta and Bodmer in $\sigma$ meson to minimize the value of $K$. With the added coupling terms, a number of parameter sets like NL1 \cite{Reinhard:1989zi}, NL2 \cite{Reinhard:1989zi}, NL3 \cite{PhysRevC.55.540} are produced, which provided the results well within the range \cite{Gambhir:1989mp}. With this, the problem of incompressibility and finite nuclei was solved, but the equation of states at high density region were quite stiff and the mass-radius of neutron stars were quite high. The addition of vector meson self coupling allowed the formation of new parameter sets \cite{BODMER1991703,PhysRevC.68.054318,GMUCA1992447,SUGAHARA1994557}, which explained both finite nuclei and infinite nuclear matter properties with a greater accuracy. \par 
The contribution of isoscalar and isovector cross couplings with new parameter sets FSUGold \cite{PhysRevLett.95.122501} and IU-FSU \cite{PhysRevC.82.055803} etc has a huge effect on neutron star radius without compromising the predictive power of finite nuclei. The introduction of $\delta$  meson \cite{KUBIS1997191,PhysRevC.89.044001} influences various quantities like symmetry energy, neutron skin thickness, neutron-proton effective masses. While the effect of $\delta$ meson on the properties of finite nuclei are minimal due to low isospin asymmetry, its contribution to the strongly isospin asymmetry matter at high densities like neutron stars is large and hence the contribution of $\delta$ meson should be considered \cite{Biswal:2016zcg}.  The inculsion of cross-couplings have a huge impact on neutron-skin thickness, symmetry energy and radius of neutron star, hence a systematic formalism based on naturalness and Naive Dimensional Analysis (NDA), the effective field theory motivated relativistic-mean-field (E-RMF) lagrangian is constructed. \par 
The E-RMF Lagrangian with exchange mesons ($\sigma,\omega$ and $\rho$) as well as $\delta$ meson and all other coupling terms is given as \cite{FURNSTAHL1996539,Kumara:2017bti,PhysRevC.97.045806}

	\begin{eqnarray}
	\fl \mathcal{E} (r) =\sum_{\alpha} \phi^{\dagger}_{\alpha} (r) \Biggl\{-i \alpha .\nabla +\beta [M-\Phi(r) -\tau_3 D(r)]+W(r) +\frac{1}{2}\tau_3 R(r)\nonumber \\
	+ \frac{1+\tau_3}{2} A(r)-\frac{i\beta \alpha}{2M}.\Bigg(f_{\omega}\nabla W(r)
	+\frac{1}{2}f_{\rho}\tau_3\nabla R(r)\Bigg)\Biggr\} \phi_{\alpha}(r)\nonumber \\
	+\Bigg(\frac{1}{2}+\frac{k_3}{3!}\frac{\Phi(r)}{M}+\frac{k_4}{4!}\frac{\Phi^2(r)}{M^2}\Bigg) \frac{m_s^2}{g_s^2}\Phi^2 (r)
	-\frac{\zeta_0}{4!}\frac{1}{g_{\omega}^2}W^4(r)\nonumber \\ 
	+\frac{1}{2 g_s^2}\Bigg(1+\alpha_1 \frac{\Phi(r)}{M}\Bigg)(\nabla \Phi(r))^2 -\frac{1}{2 g_{\omega}^2}\Bigg(1+\alpha_2 \frac{\Phi(r)}{M}\Bigg) \times (\nabla W(r))^2\nonumber \\
	-\frac{1}{2} \Bigg(1+\eta_1\frac{\Phi(r)}{M}+\frac{\eta_2}{2}\frac{\Phi^2(r)}{M^2}\Bigg)\frac{m_{\omega}^2}{g_{\omega}^2} W^2(r)-\frac{1}{2 e^2}(\nabla A(r))^2\nonumber \\ -\frac{1}{g_{\rho}^2}(\nabla R(r))^2 
	-\frac{1}{2}\Bigg(1+\eta_{\rho}\frac{\Phi(r)}{M}\Bigg)\frac{m_{\rho}^2}{g_{\rho}^2} R^2 (r) -\Lambda_{\omega}(R^2 (r) W^2(r))\nonumber \\
	+\frac{1}{2 g_{\delta}^2}(\nabla D(r))^2
	+ \frac{1}{2}\frac{m_{\delta}^2}{g_{\delta}^2} (D(r)^2),
	\end{eqnarray}\\
where $\Phi, W,R,D$ and $A$ are $\sigma,\omega,\rho,\delta$ and photon fields respectively, $g_{\sigma},g_{\omega},g_{\rho},g_{\delta}$ and $\frac{e^2}{4\pi}$ are the corresponding coupling constants and $m_{\sigma},m_{\omega},m_{\rho}$ and $m_{\delta}$ are the masses for  $\sigma, \omega, \rho$ and $\delta$ mesons respectively. $\phi_{\alpha}$ is the nucleonic field. The addition of parameters like $\eta_1$, $\eta_2$, $\eta_{\rho}$, $\alpha_1$, $\alpha_2$ in G3 set have their own importance in explaining various properties of finite as well as infinite nuclear matter. For example, the non linear interaction of $\eta_1$ and $\eta_2$ parameters analyze the surface properties of finite nuclei \cite{PhysRevC.63.024314}.\par
Using the equation $\Big(\frac{\partial \mathcal{E}}{\partial \phi_i}\Big)_{\rho = const} =0$, we obtain equation of motion for mesons. The energy-momentum tensor given by the expression\\
\begin{equation}
T_{\mu \nu} = \sum_{i} \partial_{\nu}\phi_i \frac{\partial \mathcal {L}}{\partial (\partial ^{\mu} \phi_i)} - g_{\mu \nu} \mathcal{L},
\end{equation}

gives energy density and pressure for the hadronic phase as\\
	\begin{eqnarray} \label{eq3}
	\fl \mathcal{E}_H=\sum_{i=n,p} \frac{2}{(2\pi)^3}\int_0^{k_i} d^3k E^*_i (k)+\rho W+\frac{m_s^2 \Phi^2}{g_s^2}\Bigg(\frac{1}{2}+\frac{k_3}{3!}\frac{\Phi}{M}+\frac{k_4}{4!}\frac{\Phi^2}{M^2}\Bigg)-\frac{1}{4!}\frac{\zeta_0 W^4}{g_{\omega}^2}\nonumber \\
	+\frac{1}{2}\rho_3 R -\frac{1}{2}m_{\omega^2}\frac{W^2}{g_{\omega}^2}\Bigg(1+\eta_1\frac{\Phi}{M}+\frac{\eta_2}{2}\frac{\Phi^2}{M^2}\Bigg)
	-\frac{1}{2}\Bigg(1+\frac{\eta_{\rho}\Phi}{M}\Bigg)\frac{m_{\rho}^2}{g_{\rho}^2}R^2 \nonumber \\
	 -\Lambda_{\omega}(R^2 W^2)
	+\frac{1}{2}\frac{m_{\delta}^2}{g_{\delta}^2}(D^2),
	\end{eqnarray}
and\\
\\
	\begin{eqnarray}\label{eq4}
	\fl P_H = \sum_{i=n,p}\frac{2}{3(2\pi)^3}\int_0^{k_i} d^3k E^*_i (k)-\frac{m_s^2 \Phi^2}{g_s^2}\Bigg(\frac{1}{2}+\frac{k_3}{3!}\frac{\Phi}{M}+\frac{k_4}{4!}\frac{\Phi^2}{M^2}\Bigg) 
	+\frac{1}{4!}\frac{\zeta_0 W^4}{g_{\omega}^2}\nonumber \\
	+\frac{1}{2}m_{\omega^2}\frac{W^2}{g_{\omega}^2}\Bigg(1+\eta_1\frac{\Phi}{M}+\frac{\eta_2}{2}\frac{\Phi^2}{M^2}\Bigg)
	+\frac{1}{2}\Bigg(1+\frac{\eta_{\rho}\Phi}{M}\Bigg)\frac{m_{\rho}^2}{g_{\rho}^2}R^2\nonumber \\
	 +\Lambda_{\omega}(R^2 W^2)
	-\frac{1}{2}\frac{m_{\delta}^2}{g_{\delta}^2}(D^2),
	\end{eqnarray}
where, \\
\\
$E^*_i(k) =\sqrt{k^2+ M^{*2}_i}$ is the effective energy of nucleons, $k$ is the momentum, $M_p^*$ and $M_n^*$ are the effective masses of proton and neutron  which are splitted due to $\delta$ meson\\
\begin{equation}
M_p^* = M-\Phi(r)-D(r),
\end{equation}
and
\begin{equation}
M_n^*=M-\Phi(r)+D(r).
\end{equation}
\vspace{0.2cm}
For neutron star matter, where the baryons are strongly interacting particles, the $\beta$-equilibrium and charge neutrality are two important conditions to be satisfied to determine the composition of the system. For any baryon $B$, the relation $\mu_B = b_B \mu_n - q_B \mu_e$, where $\mu_B$ is the chemical potential with charge $q_B$ and baryon number $b_B$, represents the beta-equilibrium condition. For the present case with $n,p$ and $e$ only, the $\beta$-equilibrium condition is given by the chemical potential of proton $\mu_p$, neutron $\mu_n$ and electron $\mu_e$ as
\begin{equation}\label{c1}
\mu_p = \mu_n - \mu_e. 
\end{equation}

The chemical potential of a baryon can thus be obtained from these two independent chemcial potentials $\mu_n$ and $\mu_e$. The charge neutrality condition is given by\\
\begin{equation}
q_{total} = \sum_{i=n,p} q_i k_i^3/(3\pi^2)+\sum_l q_l k_l^3/(3\pi^2)=0,
\end{equation}
\vspace{0.2cm}
which implies, $n_p$ = $n_e$, where $n_p$ and $n_e$ are the number densities of proton and electron respectively. \\
The total energy density and pressure of neutron star matter is then given as\\
\begin{equation*}
\mathcal{E} = \mathcal{E}_H +\mathcal{E}_l,
\end{equation*}
\begin{equation}
P=P_H+P_l
\end{equation}
$\mathcal{E}_l$ and $P_l$ are the lepton energy density and pressure.
\begin{equation}
\mathcal{E}_l= \sum_{l=e} \frac{2}{(2\pi^3)}\int_0^{k_l} d^3k \sqrt{k^2 +m_l^2},
\end{equation}
and\\
\begin{equation}
P_l= \sum_{l=e} \frac{2}{3(2\pi^3)}\int_0^{k_l} d^3k k^2 /(\sqrt{k^2 +m_l^2})
\end{equation}

\subsection*{B. Quark Matter}
The density in the central part of the neutron star is presumed to be high enough for the hadron matter to undergo a phase transition to quark matter. This transition leads to the formation of a mixed phase at the density that varies from saturation density $\rho_0$ to few times $\rho_0$ depending upon the properties of NS and the models used. For the quark phase, we employ the simple MIT Bag model for the unpaired quark matter \cite{PhysRevD.9.3471,PhysRevD.17.1109,PhysRevD.30.2379}. This model is a degenerate Fermi gas of quarks (u,d and s) and electrons with chemical equilibrium being maintained by several weak interactions. In this model, the quarks are assumed to be confined in a colorless region where the quarks are free to move. The quark masses considered are as $m_u$= $m_d$ =5.0 MeV and $m_s$ = 150 MeV. For the present work, we ignore the one gluon exchange inside the gas.The equilibrium condition satisfied by the quark matter is\\
\begin{equation}\label{c2}
\mu_d=\mu_s=\mu_u + \mu_e.
\end{equation}
\vspace{0.2cm}
The chemical potential of the individual quark follows from the neutron and electron chemical potentials $\mu_n$ and $\mu_e$ respectively as:\\
\begin{equation}
\mu_u =\frac{1}{3}\mu_n -\frac{2}{3}\mu_e,
\end{equation}
\begin{equation}
\mu_d =\frac{1}{3}\mu_n +\frac{1}{3}\mu_e,
\end{equation}
and
\begin{equation}
\mu_s =\frac{1}{3}\mu_n +\frac{1}{3}\mu_e.
\end{equation}
\vspace{0.2cm}
The charge neutrality condition obtained is\\
\begin{equation}
\frac{2}{3}n_u-\frac{1}{3}n_d -\frac{1}{3}n_s -n_e=0,
\end{equation}
\vspace{0.2cm}
where, $n_q (q=u,d,s,e)$. The total quark matter density is given as\\
\begin{equation}
n_Q = \frac{1}{3}(n_u +n_d +n_s).
\end{equation}

The pressure of the quarks (q=u,d,s) is given by \cite{kapusta_gale_2006}\\
\begin{equation}
P_Q = \frac{1}{4\pi^2}\sum_q \Biggl\{\mu_q k_q \Bigg(\mu_q^2 -\frac{5}{2}m_q^2\Bigg)+\frac{3}{2}m_q^4 ln \Bigg(\frac{\mu_q +k_q}{m_q}\Bigg)\Biggr\}.
\end{equation}
\vspace{0.2cm}

The total pressure due to quarks and leptons is given by
\begin{equation}\label{eq16}
P=P_Q +P_l -B,
\end{equation}

where, $B$ is the Bag constant. The bag constant is the difference in the energy densities of the perturbative vacuum and the non-perturbative vacuum (true ground state of QCD). The pressure exerted by the freely moving quarks at the surface of the bag can make the bag unstable. To prevent this an external pressure defined as the Bag pressure $B$ is applied to compensate the internal pressure of the system.\par 

The expression for the quark energy density is\\ 
\begin{equation}\label{eq17}
\mathcal{E}_Q = \frac{3}{4\pi^2}\sum_q \Biggl\{\mu_q k_q \Bigg(\mu_q^2 -\frac{1}{2}m_q^2\Bigg)-\frac{1}{2}m_q^4 ln \Bigg(\frac{\mu_q +k_q}{m_q}\Bigg)\Biggr\}+ B
\end{equation}  			
\vspace{0.2cm}
A range of bag constants have been used in the literature\cite{Baym:2017whm,STEINER2000239,BUBALLA2005205,NOVIKOV1981301}. In the bag model, the standard value of $B$ is taken as $B^{1/4}$= 140 MeV \cite{PhysRevD.22.1198, PhysRevD.12.2060}  In our previous work \cite{rather2020constraining}, we have constrained the value of bag constant for hybrid stars. Considering the range varying from $B^{1/4}$= 100-200 MeV, we found that the bag values 130 MeV$<$$B^{1/4}$$<$ 160 MeV are  suitable for explaining the presence of quark matter phase in neutron stars.\par 

\subsection*{C: Mixed Phase}
The deconfined phase transition from hadron matter to quark matter is assumed to be of first order, so the transition should produce a mixed phase between the pure hadron phase and pure quark phase. The mixed phase region between the pure hadron matter and the quark matter is not well defined \cite{PhysRevD.46.1274}. Beta-equilibrium and charge neutrality conditions determine the density range over which the mixed phase can exist. The quark-hadron phase transition in neutron stars has been widely studied using different techniques \cite{PhysRevD.46.1274,PhysRevC.60.025801,PhysRevC.75.035808,PhysRevC.66.025802,PhysRevC.89.015806}. Usually the technique involved in constructing the mixed phase depends upon the surface tension. Beyond a critical value of the surface tension, the Maxwell construction (MC) \cite{PhysRevD.88.063001} is used. With no specific value of surface tension being known, the Gibbs construction (GC) \cite{PhysRevD.46.1274} is found to be more relevant.
The MC is appropriate to obtain the liquid-vapor phase transition EoS. However, Glendenning \cite{PhysRevD.46.1274,GLENDENNING2001393} pointed out that MC is not appropriate for the hadron-quark phase transition. Glendenning also pointed out that the usual Maxwell construction is applicable for systems with one particle species and correspondingly one chemical potential, whereas in neutron stars there are two relevant quantities, the charge and baryon number chemical potentials. In GC, the global charge neutrality is imposed which means that both hadron phase and quark phase are allowed to be charge neutral separately, whereas in Maxwell construction, local charge neutrality condition is used. Also, in GC, the pressure increases with the density in the mixed phase contrary to Maxwell construction, where the pressure remains constant throughout the phase transition.\par 
The Gibbs conditions for the mixed phase are given by:\par 

\begin{equation}\label{d1}
P_{HP}(\mu_{HP}) = P_{QP}(\mu_{QP}) = P_{MP},
\end{equation}
and
\begin{equation}\label{d2}
\mu_{HP,i} = \mu_{QP,i} = \mu_i, ~i=n,e.
\end{equation}
\vspace{0.2cm}
In case of two independent chemical potentials which follow from eqs.(\ref{c1}) and (\ref{c2}), the gibbs conditions (eqs.\ref{d1},\ref{d2}) can be fulfilled if the coexisting phases have opposite electric charges with global charge neutrality imposed, the baryon density for the mixed phases then follows from the equation:\par 
\begin{equation}\label{e2}
\rho_{MP} = \chi \rho_{QP} +(1-\chi)\rho_{HP}.
\end{equation}
\vspace{0.2cm}
where, $\chi$ = $V_Q/V$ is the volume fraction of the quark phase.  The quark volume fraction $\chi$ is obtained using the global charge neutrality of the mixed phase within the volume $V$ which implies that the charge density integral $Q=4\pi \int_V dr r^2 q(r)$, must vanish rather than $q(r)$ itself.
\begin{equation}
0=\frac{Q}{V} = (1-\chi)q_H (\mu^n, \mu^e)+\chi q_Q (\mu^n, \mu^e) +q_L
\end{equation}
where, $q_L$ is the lepton charge density. The energy density in the mixed phase then reads: 
\begin{equation}\label{e1}
\varepsilon_{MP} = \chi \varepsilon_{QP} +(1-\chi)\varepsilon_{HP} +\varepsilon_l,
\end{equation}
The $\chi$, by definition, varies between 0 and 1 depending on how much the hadronic matter has been converted to the quark matter.\\

Once the mixed phase is obtained, the eqs.(\ref{e1}) and (\ref{e2}) can be solved to determine the properties of the mixed phase.

\section{Symmetry Energy}
The symmetry energy $S$ for a nuclear system with mass number $A$ is defined as $S =\frac{E}{A}(A,N=A)-\frac{E}{A}(A,N= Z)$. Huge literature is devoted to the calculation of the symmetry energy $S$ and its slope parameter $L$. Different phenomenological approaches like Hartree-Fock \cite{FARINE1978317} and Thomas-Fermi \cite{PEARSON19911} have been used to study the symmetry energy which predict the value of symmetry energy in the range 27-38 MeV at saturation. Such studies have also shown the correlation between the slope parameter and the neutron skin thickness.\par 

The energy density $\mathcal{E}(\rho,\alpha)$ can be approximated by the parabolic law as \cite{PhysRevC.44.1892}
\begin{equation}
\mathcal{E}(\rho, \alpha)= \mathcal{E}(\rho) + S(\rho) \alpha^2 + \mathcal{O}(\alpha^4),
\end{equation}
\vspace{0.2cm}
where, $\mathcal{E}(\rho)$ is the energy density of symmetric nuclear matter ($\alpha=0$) and $S(\rho)$ is the symmetry energy defined as
\begin{equation}
S(\rho) = \frac{1}{2}\Big[\frac{\partial^2 \mathcal{E}(\rho,\alpha)}{\partial \alpha^2}\Big]_{\alpha=0}.
\end{equation}
\vspace{0.2cm}
This isospin asymmetry arise as a result of difference in the masses and densities of proton and neutron. The isovector-vector meson $\rho$ takes care of asymmetry density while the isovector-scalar meson $\delta$ takes care of mass asymmetry. The combined expression of the $\rho$ and $\delta$ meson symmetry energies gives the overall symmetry energy of the system  \cite{PhysRevC.63.024314,KUBIS1997191,PhysRevC.84.054309}
\begin{equation}
S(\rho) = S^{kin}(\rho)+ S^{\rho}(\rho) +S^{\delta}(\rho),
\end{equation}
where, 
\begin{equation}
S^{kin}(\rho) =\frac{k_F^2}{6 E_F^*}
\end{equation}
and
\begin{equation}
S^{\rho}(\rho) =\frac{g_{\rho}^2 \rho}{8 m_{\rho}^{*2}}.
\end{equation}
\vspace{0.2cm}
Due to the cross-coupling between the $\rho$-$\omega$ fields, the mass of the $\rho$ meson is modified as
\begin{equation}
m_{\rho}^{*2} = \Big(1+\eta_{\rho} \frac{\Phi}{M}\Big)m_{\rho}^2 +2 g_{\rho}^2 (\Lambda_{\omega} W^2).
\end{equation}
\vspace{0.2cm}
The contribution to the symmetry energy due to the $\delta$ meson is\\
\begin{equation}
S^{\delta}(\rho) =-\frac{1}{2}\rho \frac{g_{\delta}^2}{m_{\delta}^2} \Bigg(\frac{M^*}{E_F}\Bigg)^2 u_{\delta}(\rho, M^*).
\end{equation}
\vspace{0.2cm}
The function $u_{\delta}$ follows from the discreteness of the Fermi momentum. In nuclear matter, this momentum is quite large and hence the system can be treated to be continuous which implies that the function $u_{\delta} \approx $1. So the final expression for the symmetry energy becomes\\
\begin{equation}
S(\rho) = \frac{k_F^2}{6 E_F^*}+\frac{g_{\rho}^2 \rho}{8 m_{\rho}^{*2}} -\frac{1}{2}\rho \frac{g_{\delta}^2}{m_{\delta}^2} \Bigg(\frac{M^*}{E_F}\Bigg)^2.
\end{equation}
\vspace{0.2cm}
Numerically, the symmetry energy $S(\rho)$ is calculated as the difference in the energy of the Symmetric Nuclear Matter (SNM) and Pure Neutron Matter (PNM). The symmetry energy around the saturation density $\rho_0$ can be expanded by Taylor series as:\\
\begin{equation}
S(\rho) = J+L \mathcal{Y}+ \frac{1}{2} K_{sym} \mathcal{Y}^2 + \frac{1}{6} Q_{sym} \mathcal{Y}^3 +\mathcal{O}[\mathcal{Y}^4],
\end{equation}
\vspace{0.2cm}
where, \par 
$J$=$S(\rho_0)$ is the symmetry energy at the saturation density $\rho_0$ and $\mathcal{Y} = (\rho- \rho_0)/(3\rho_0)$. The derivatives of $S(\rho)$ are $L$, $K_{sym}$ and $Q_{sym}$ and are defined as:\\
\begin{equation}
L = 3 \rho_0 \frac{\partial S(\rho)}{\partial \rho}\Bigg|_{\rho = \rho_0},
\end{equation}
\vspace{0.2cm}
\begin{equation}
K_{sym} = 9 \rho_0^2 \frac{\partial^2 S(\rho)}{\partial \rho^2}\Bigg|_{\rho = \rho_0},
\end{equation}
\vspace{0.2cm}
and\\
\begin{equation}
Q_{sym} = 27 \rho_0^3 \frac{\partial^3 S(\rho)}{\partial \rho^3}\Bigg|_{\rho = \rho_0}.
\end{equation}
\vspace{0.2cm}
Here, $L$ is the slope parameter and $K_{sym}$ represents the symmetry energy curvature at saturation density. $Q_{sym}$ is the skewness of $S(\rho)$ at $\rho_0$. To fix the values of all these quantities, a large number of attempts have been made \cite{Singh_2013,PhysRevC.86.015803,PhysRevC.85.035201,PhysRevC.82.054607,Newton_2012,PhysRevLett.108.081102,PhysRevC.86.025804}.The density dependent symmetry energy is an important quantity to understand the properties of both finite as well as infinite matter \cite{PhysRevLett.102.122502}. The currently accepted values of symmetry energy and its slope are $J$ = 31.6 $\pm$ 2.66 MeV and $L$ = 58.9$\pm$ 16 MeV, which are obtained from various astrophysical observations \cite{LI2013276}. The precise values of these quantities are yet to be determined experimentally.\par 
The symmetry energy and its density dependence have a strong correlation between the pressure (at $\rho \approx \rho_0$) inside a neutron star and its radius \cite{Lattimer_2001}. Also, studies have shown that the slope parameter $L$ is related to the neutron skin thickness. A large value of $L$ corresponds to a higher  neutron matter pressure and a thicker neutron skin \cite{FURNSTAHL200285,PhysRevLett.106.252501,PhysRevLett.86.5647}. It is found that the value of the parameters $L$, $K_{sym}$ and $Q_{sym}$ have a huge impact on the Radius-Mass relation of a neutron star \cite{PhysRevC.97.025806}. The more accurate values of these parameters may come from the future experiments or from a better knowledge of neutron star MR relation. \par 
For hadron EoS, all these quantities are known with some uncertainities. But for hybrid EoS, no such measurement has been made. In this work, we calculate all these quantities for a hybrid EoS which heavily influence the neutron star MR relation.

\section{Results and Discussions}
To calculate the symmetry energy and all other parameters for a hybrid EoS, we used different parameter sets NL3 \cite{PhysRevC.55.540}, FSUGarnet \cite{CHEN2015284}, G3 \cite{KUMAR2017197} and IOPB-I \cite{PhysRevC.97.045806} for hadron matter. The NM  properties for the hadron EoS at saturation density $J$ , $L$, $K_{sym}$ and $Q_{sym}$ for all parameter sets are listed in table \ref{tab1}. For NL3 set, the symmetry energy $J$ = 37.43 MeV and slope parameter $L$ = 118.65 MeV are little higher than the empirical value $J$ = 31.6 $\pm$ 2.66 MeV and $L$ = 58.9 $\pm$ 16 MeV  \cite{LI2013276}. The  $J$ and $L$ for other parameter sets lie well within the given range. $K$ is the nuclear matter incompressibility at saturation that determines the extent to which a nuclear matter can be compressed and is defined as \\
\begin{equation}
K = 9\rho_0 \frac{\partial^2 \mathcal{E}}{\partial \rho^2}\Bigg|_{\rho = \rho_0}.
\end{equation}
\vspace{0.2cm}
The current accepted value of $K$=240 $\pm$ 20 MeV is determined from the isoscalar giant monopole resonance (ISGMR) for $^{90}Zr$ and $^{208}Pb$ \cite{Colo:2013yta,Piekarewicz:2013bea}. The incompressibility of the given parameter sets lie within the range 240$\pm$20 MeV with NL3 set producing a little higher value than the rest. The G3 set predicts more accurate value of $K$=243.96 MeV, which shows that the contribution of $\delta$ mesons in necessary for high dense matter.\par 
\begin{table}[ht]
	\centering
	\caption{\label{tab1} Parameter sets and the corresponding nuclear matter properties for hadron matter. For all the sets, the nucleon mass is $M$= 939.0 MeV. All the coupling constants are dimensionless except $k_3$ which has the dimensions of fm$^{-1}$. }
	\begin{indented}
		\item[] \begin{tabular}{ ccccc }
		\br
		&NL3&FSUGarnet&G3&IOPB-1 \\
		\mr
		$m_s/M$ & 0.541&0.529&0.559&0.533\\
		$m_{\omega}/M$ &0.833&0.833&0.832&0.833\\
		$m_{\rho}/M$&0.812&0.812&0.820&0.812\\
		$m_{\delta}/M$&0.0&0.0&1.043&0.0\\
		$g_s/{4\pi}$&0.813&0.837&0.782&0.827\\
		$g_{\omega}/{4\pi}$&1.024&1.091&0.923&1.062\\
		$g_{\rho}/{4\pi}$&0.712&1.105&0.962&0.885\\
		$g_{\delta}/{4\pi}$&0.0&0.0&0.160&0.0\\
		$k_3$&1.465&1.368&2.606&1.496\\
		$k_4$&-5.688&-1.397&1.694&-2.932\\
		$\zeta_0$&0.0&4.410&1.010&3.103\\
		$\eta_1$&0.0&0.0&0.424&0.0\\
		$\eta_2$&0.0&0.0&0.114&0.0\\
		$\eta_{\rho}$&0.0&0.0&0.645&0.0\\
		$\Lambda_{\omega}$&0.0&0.043&0.038&0.024\\
		$\alpha_1$&0.0&0.0&2.000&0.0\\
		$\alpha_2$&0.0&0.0&-1.468&0.0\\
		$f_{\omega}/4$&0.0&0.0&0.220&0.0\\
		$f_{\rho}/4$&0.0&0.0&1.239&0.0\\
		$\beta_{\sigma}$&0.0&0.0&-0.087&0.0\\
		$\beta_{\omega}$&0.0&0.0&-0.484&0.0\\
		\mr
		\mr
		$\rho_0$ (fm$^{-3}$) & 0.148 &0.153&0.148&0.149\\
		$\epsilon_0 $(MeV) & -16.29&16.23&-16.02&-16.10 \\
		M*/M&0.595&0.578&0.699&0.593\\
		$J $(MeV)  & 37.43&30.95&31.84&33.30 \\
		$L $(MeV) &118.65&51.04&49.31&63.58 \\
		$K_{sym}$ (MeV) &101.34&59.36&-106.07&-37.09 \\
		$Q_{sym}$ (MeV)& 177.90&130.93&915.47&862.70 \\
		$K$ (MeV)&271.38&229.5&243.96&222.65\\
		\br
	\end{tabular}
\end{indented}
\end{table}
The value of incompressibility for different parameter sets are compatible with the observational data from various experiments. The value of incompressibility parameter at saturation density is an important feature of nuclear matter. It appears as a parameter in the calculations of mass spectrum and properties of neutron stars, which are important in understanding the nuclear matter at high densities.\par 

\begin{figure}
	\centering
	\includegraphics[width=0.6\textwidth]{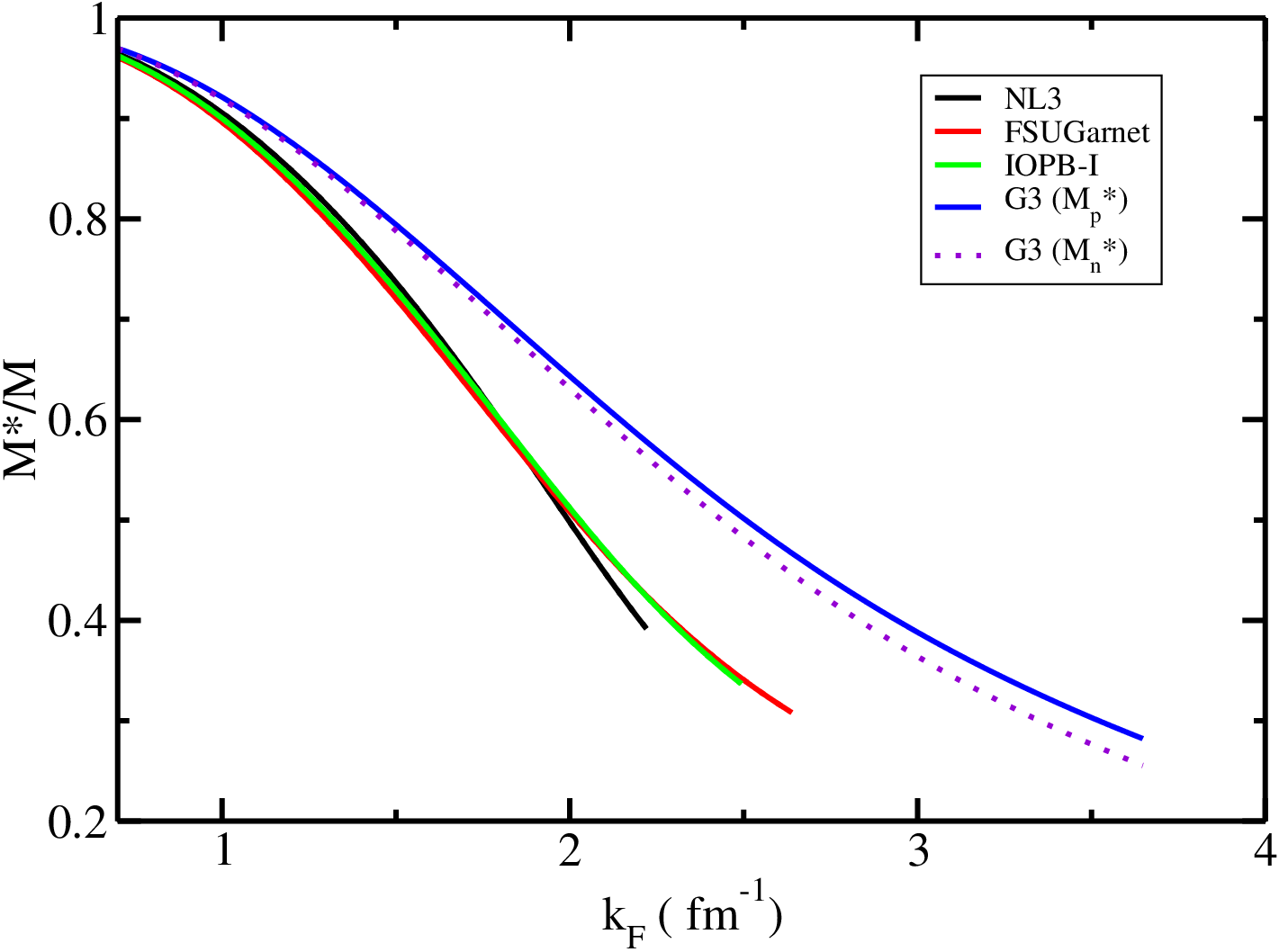}
	\caption{Effective masses of proton and neutron as a function of Fermi momentum for different E-RMF parameters.}
	\label{fig1}
\end{figure}

To obtain energy density and the pressure for neutron star matter in $\beta$-equilibrium and charge neutrality condition, we solve equation (\ref{eq3}) and equation (\ref{eq4}) for different parameter sets. Figure (\ref{fig1}) shows the variation of proton and neutron effective mass as a function of Fermi momentum for different  parameter sets. For G3 set, the effective masses of proton $M_p^*$ and neutron $M_n^*$ are different due to the the contribution from the $\delta$ meson. At very low momentum, both $M_p^*$ and $M_n^*$ overlap each other, but as the momentum increases, the effective masses of proton and neutron split. The solid (blue) line represents the proton effective mass and the dotted (violet) line corresponds to the effective mass of neutron for G3 set. For NL3, FSUGarnet and IOPB-I sets, the proton and neutron effective masses overlap as there is no contribution from $\delta$ meson. Physical properties like symmetry energy, neutron-skin thickness, isotopic shift are effected due to the $\delta$ meson inculsion. Thus, it is important to include the contribution of $\delta$ meson. \par 

\begin{figure}[h]
	\centering
	\includegraphics[width=0.6\textwidth]{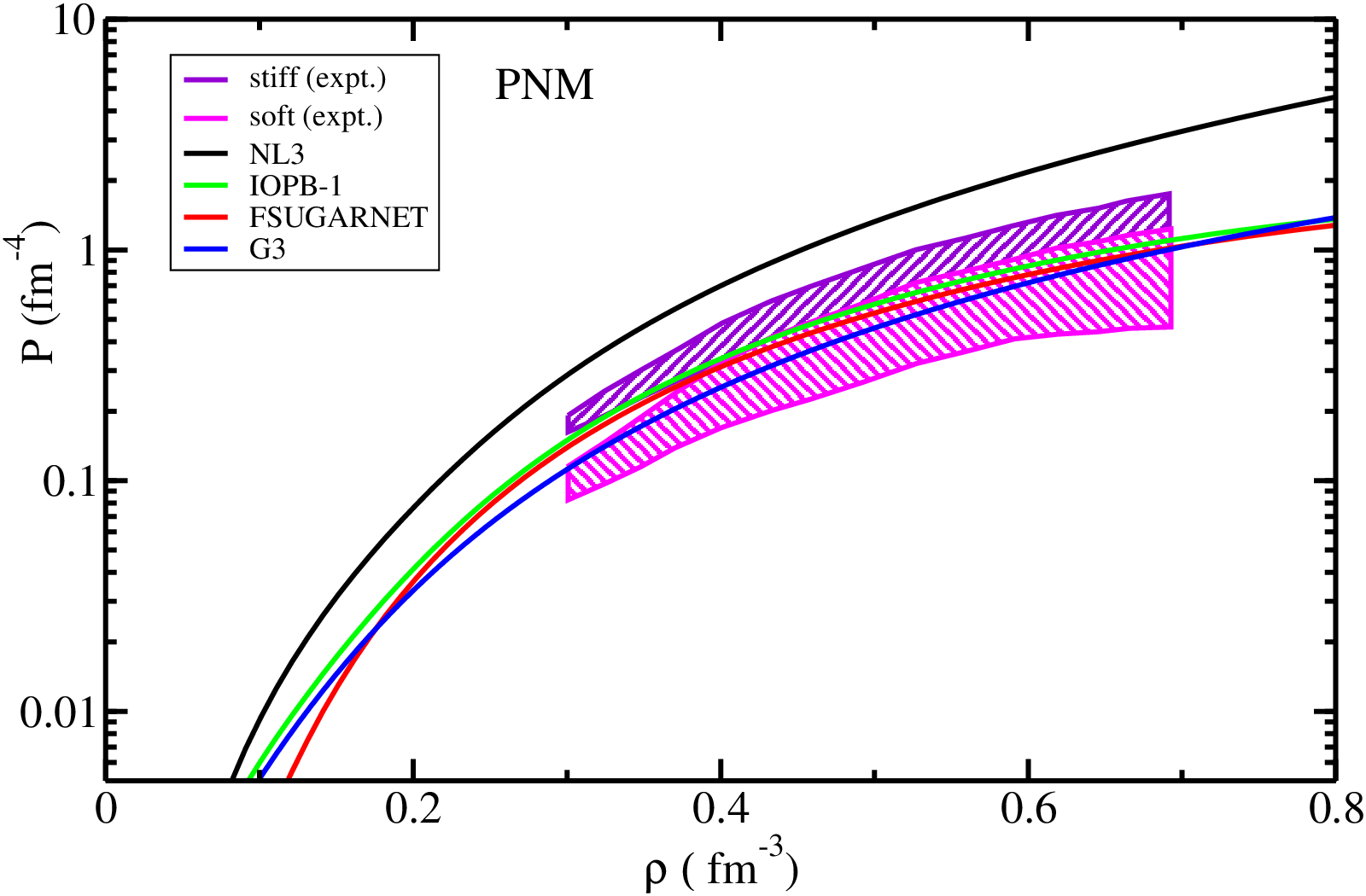}
	\caption{Pressure vs baryon density for Pure Neutron Matter (PNM)  with different E-RMF parameters.}
	\label{fig2}
\end{figure}

 Figure (\ref{fig2}) shows the variation of pressure P with the baryon density for pure neutron matter (PNM).  The results are compared with the experimental flow data obtained from the analysis of heavy ion collisions \cite{Danielewicz1592},where upper one (stiff-expt.) corresponds to the strong density dependence of $S(\rho)$ and the lower one (soft-expt.) corresponds to the weak dependence. It is clear from figure (\ref{fig2}) that the PNM EoS for G3 set is compatible with the experimental data. The NL3 set produces stiffer results than the other forces at high densities. The IOPB-I and FSUGarnet EoS is also compatible with the data.



\begin{figure}[h]
	\centering
	\includegraphics[width=0.6\textwidth]{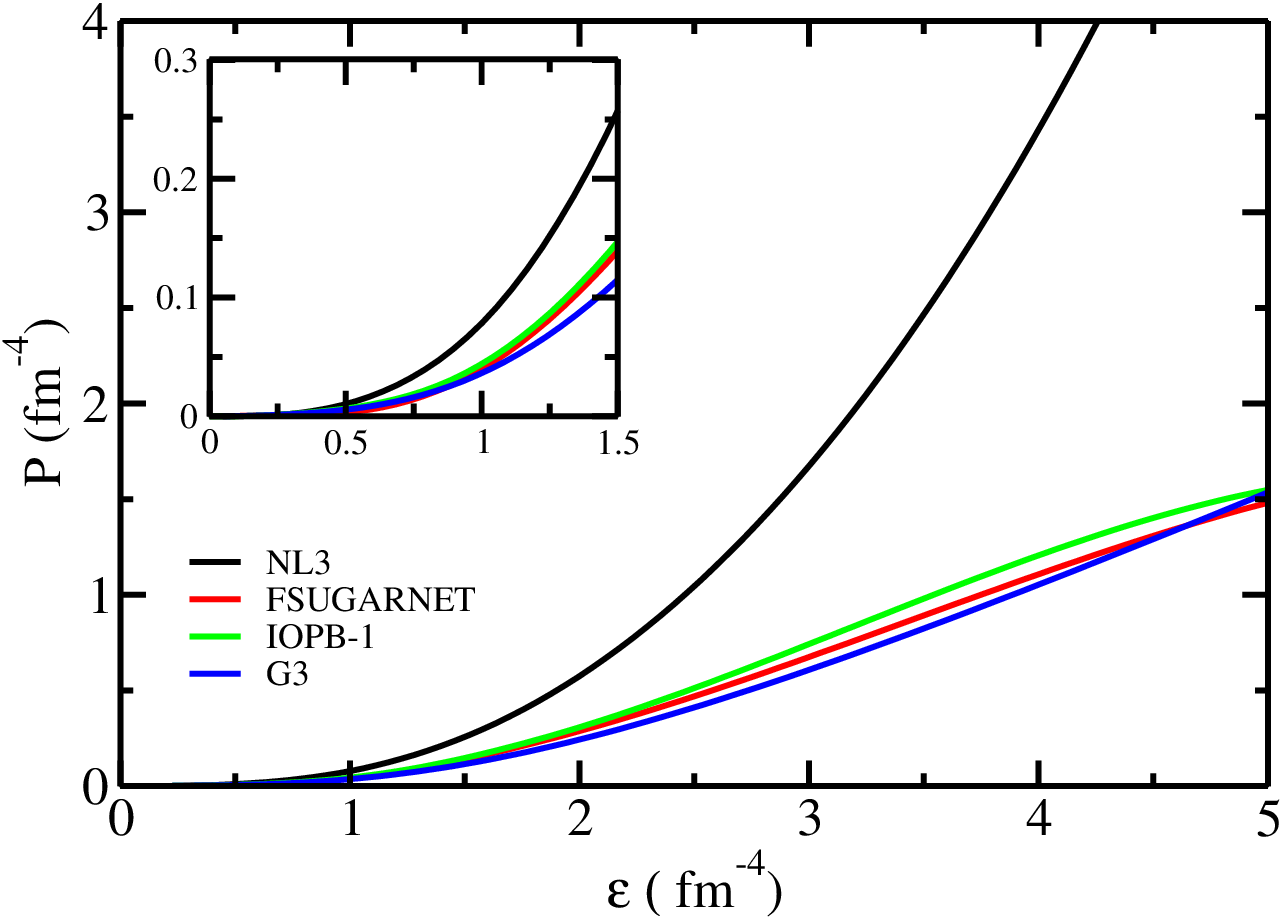}
	\caption{EoS for NS matter in $\beta$-equilibrium and charge neutrality condition with different parameter sets.}
	\label{fig3}
\end{figure}

Figure (\ref{fig3})  displays the variation of pressure with energy density for $\beta$-equilibrated charge neutral neutron star matter for parameter sets NL3, FSUGarnet, IOPB-I and G3. The NL3 parameter set yields a stiffer EoS. FSUGarnet and IOPB-I  have similar EoSs at high density but they differ slightly at low energy density. FSUGarnet has soft eos at low energy density $\mathcal{E} \approx 0.5$ fm$^{-4}$  but becomes stiff at higher density as compared to G3. The G3 set provides the soft EoS.  \\

\begin{figure}[h]
	\centering
	\includegraphics[width=0.6\textwidth]{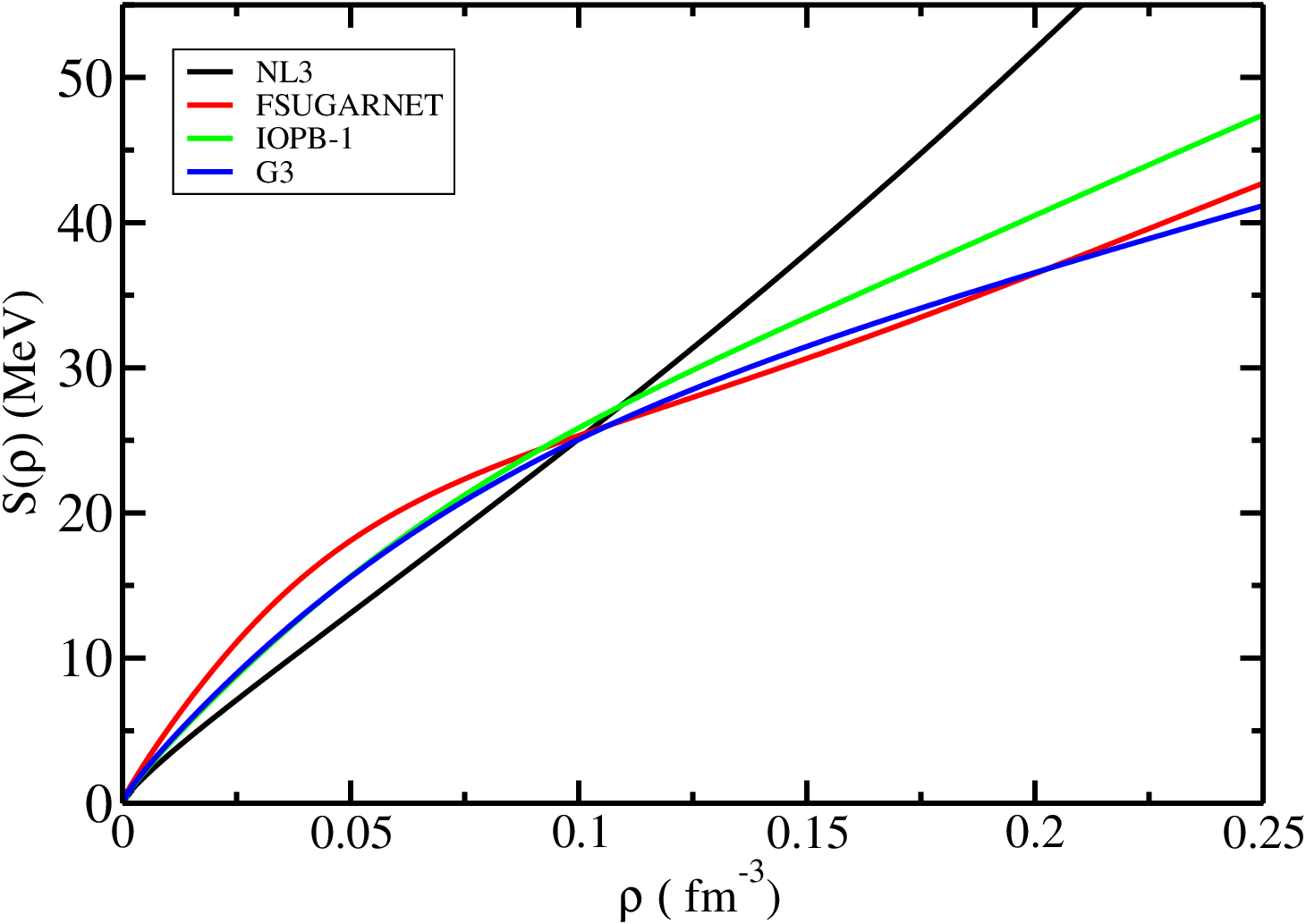}
\caption{Density dependence of symmetry energy for different E-RMF parameters.}
\label{fig4}
\end{figure}

The symmetry energy $S(\rho)$ as a function of density is displayed in figure (\ref{fig4}). The symmetry energy for NL3 set is stiff at high density as compared to FSUGarnet, IOPB-I and G3 parameter sets which provide soft $S(\rho)$. The presence of $\rho-\omega$ cross-coupling in IOPB-I and FSUGarnet sets and $\rho-\sigma$ in G3 set yields softer symmetry energy. The slope parameter $L$ and the symmetry energy curvature $K_{sym}$ are smaller in G3 as compared to the others as shown in table (\ref{tab1}). This implies that the G3 set has softer symmetry energy at high density. This effect of symmetry energy plays an important role in the cooling process of neutron star.

\begin{figure}[h]
	\centering
	\includegraphics[width=0.6\textwidth]{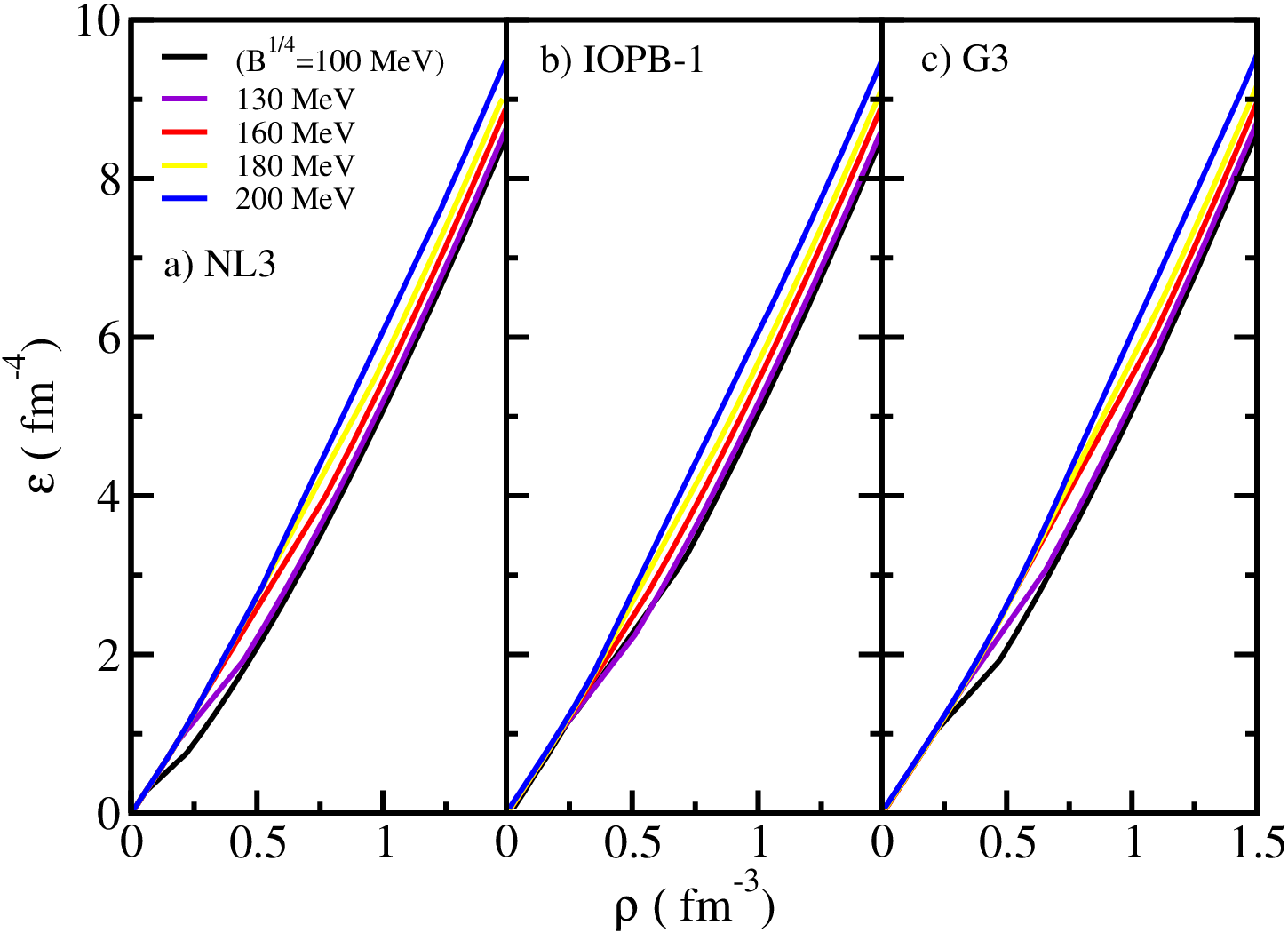}
	\caption{Hybrid EoS for  different bag constants for a) NL3, b) IOPB-I and c) G3 parameter sets.}
	\label{fig5}
\end{figure}

To obtain hybrid EoS, we solve equation ({\ref{e1}}) and equation (\ref{e2}) together with the hadronic and quark EoS.
All the hybrid EoS for different hadronic matter parameter sets (NL3, IOPB-I and G3) and for different quark matter bag values ($B^{1/4}$ = 100, 130, 160, 180 and 200 MeV) are shown in figure (\ref{fig5}).  It is clear that the energy density increases with the bag constant and hence the pressure will correspondingly decreases with the bag constant. This implies that the hybrid EoS becomes more softer as we increase the bag value. It is to be mentioned that the phase transition density of mixed phase changes with the bag constant. For small values of $B$, the phase transition takes place below the nuclear saturation density\cite{Ghosh1995}. As the bag value increases, the phase transition density shifts to higher values. The importance of hybrid EoS lies in the formation of mixed phase. The transition from HM to QM using Gibbs condition determines the stiffness or softness of the EoS. Due to the stiffness/softness of hybrid EoS by the mixed phase, the nuclear matter properties of hybrid EoS change with the bag constant. \par 
\vspace{0.2cm}
\begin{table}[ht]
	\centering
	\caption{\label{tab2} Transition densities for the mixed phase. $\rho^{MP}_{start}$ and $\rho^{MP}_{end}$ denote the formation and the end of the mixed phase respectively. }
	\begin{indented}
		\item[] \begin{tabular}{ cccccc }
			\br
	\centering
			$B^{1/4}$ (MeV)&100&130&160&180&200 \\
			\mr
			&&NL3\\
			\mr	
		$\rho^{MP}_{start}$($\rho_0$)  & 0.98&1.12&1.81&3.05&4.63 \\
		$\rho^{MP}_{end}$($\rho_0$) &1.43&2.44&3.22&6.12&8.12\\
		
			\mr
			&& IOPB-I\\
			\mr	
		$\rho^{MP}_{start}$($\rho_0$) & 0.96&1.04&1.63&2.96&4.42 \\
		$\rho^{MP}_{end}$($\rho_0$) &1.16&2.14&3.02&5.81&7.93\\
			
			\mr
			&& G3\\
			\mr	
		$\rho^{MP}_{start}$ ($\rho_0$)  & 0.96&1.02&1.58&2.84&4.16\\
		$\rho^{MP}_{end}$ ($\rho_0$) &1.20&2.06&2.92&5.34&7.15\\
			
			\br
		\end{tabular}
	\end{indented}
\end{table}
Table \ref{tab2} shows the transition densities for the mixed phase. $\rho^{MP}_{start}$ represents the end of pure hadron phase and beginning of hadron-quark mixed phase, while  $\rho^{MP}_{end}$ represents the beginning of pure quark phase. For $B^{1/4}$=100 MeV, the transition density from pure hadron phase to mixed phase occurs at around $\approx \rho_0$. With increasing bag constant, the phase transition density also increases. For $B^{1/4}$=200 MeV, the mixed phase region extends from $\approx$ (4-8)$\rho_0$. It is clear that the mixed phase region broadens with the bag constant.\\

From the EoSs obtained (figure \ref{fig5}), quantities like energy density, pressure and density are now known for the hybrid EoS. The nuclear matter properties like symmetry energy and other quantities for the hybrid EoS at saturation as are calculated as shown in table \ref{tab3}.

\begin{table}[ht]
	\centering
	\caption{\label{tab3} NM properties of Mixed EoS for different bag constants. }
	\begin{indented}
	\item[] \begin{tabular}{ cccccc }
		\br
		$B^{1/4}$ (MeV)&100&130&160&180&200 \\
		\mr
		&&NL3\\
		\mr	
		$J $(MeV)  & 45.11&41.72&35.76&32.20&36.84 \\
		$L $(MeV) &130.75&128.12&124.59&121.05&131.78\\
		$K_{sym}$ (MeV) &831.63&842.93&832.35&842.42&806.18 \\
		$Q_{sym} $(MeV)&2047.88&4004.93&5052.79&5068.79&5251.48\\
		$K$ (MeV)&580.08&566.95&557.83&554.43&522.84\\
		\mr
		&& IOPB-I\\
		\mr	
		$J $(MeV)  & 35.88&37.86&38.45&43.64&54.54 \\
		$L $(MeV) &69.61&62.28&68.64&72.18&89.32\\
		$K_{sym}$ (MeV) &419.19&429.87&432.53&472.08&496.86 \\
		$Q_{sym}$ (MeV)&2289.71&2655.78&2679.69&2796.74&2886.82\\
		$K$ (MeV) &455.76&432.61&415.29&401.08&400.85\\
		\mr
		&& G3\\
		\mr	
		$J$ (MeV)  & 37.48&37.89&38.71&51.49&56.17\\
		$L$ (MeV) &55.66&55.82&68.73&76.39&81.05\\
		$K_{sym}$ (MeV) &329.38&330.14&387.48&395.96&415.25 \\
		$Q_{sym}$ (MeV)&5693.23&5715.55&5910.85&6330.28&6421.93\\
		$K$ (MeV) & 557.03&543.93&540.79&539.58&537.24\\
		\br
	\end{tabular}
\end{indented}

\end{table} 
\vspace{0.5cm}
The value of symmetry energy $J$ at saturation and other parameters are very large as compared to the pure hadronic matter. The $J$ value of hadronic EoS for G3 set is 31.84 MeV, while for G3 hybrid EoS the value is 37.48 MeV  for $B^{1/4}$=100 MeV and increases with the bag constant. The value of slope parameter for hybrid EoS with G3 force lies in the range (50-80) MeV which is compatible with the astrophysical observations \cite{LI2013276}, but for NL3 hybrid EoS, the $L$ value is very large and lies in the range (120-130) MeV. Similarly, the value of $Q_{sym}$ for all the hybrid EoSs lies in the range (2000-6000) MeV which is quite large compared to the values obtained for pure hadronic matter (100-900 MeV). \par
\begin{figure}[h]
	\centering
	\includegraphics[width=0.6\textwidth]{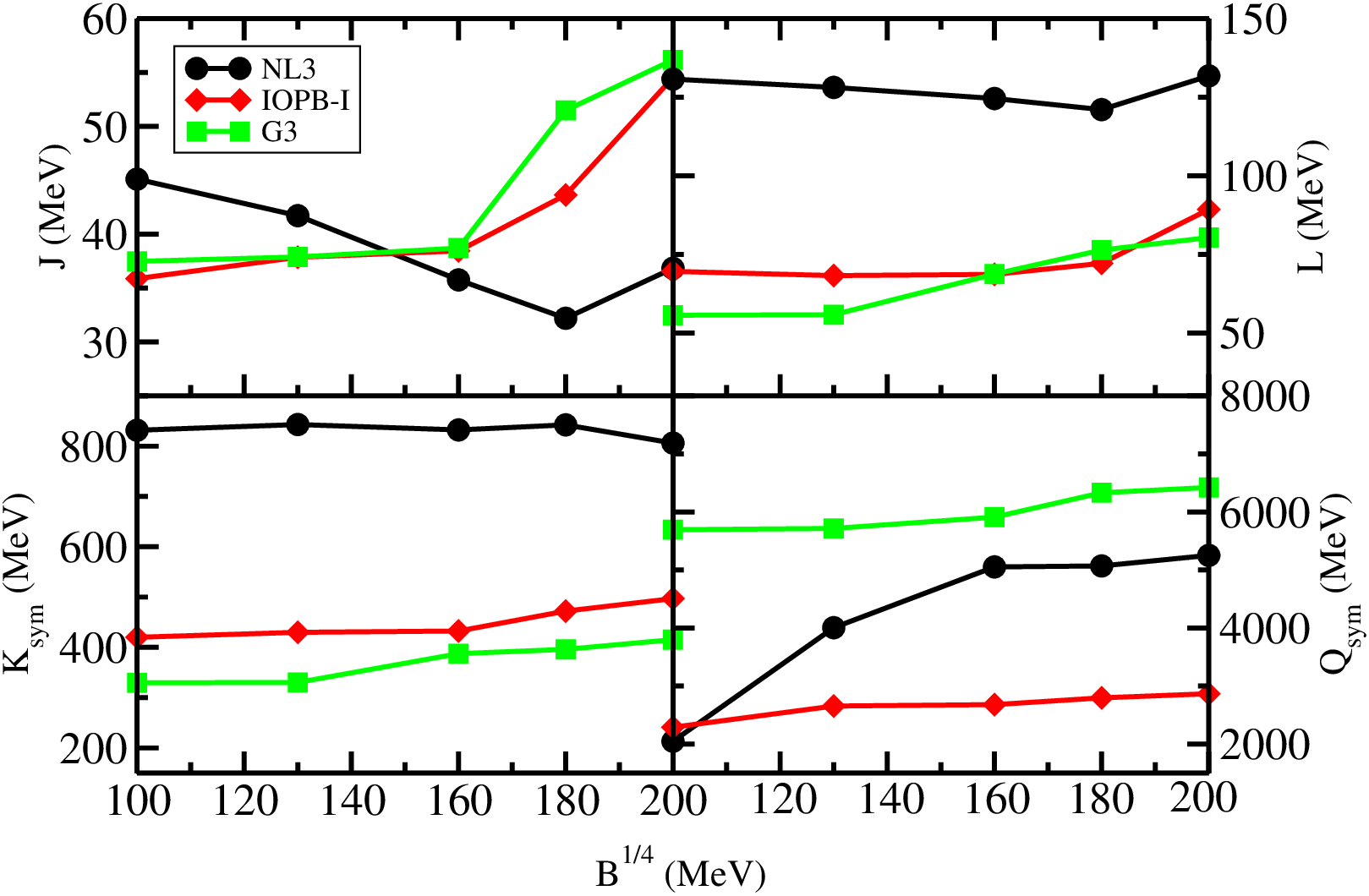}
	\caption{ NM properties for Hybrid EoS as a function of bag constant $B^{1/4}$ for a) NL3, b) IOPB-I and c) G3 parameter sets.}
	\label{fig6}
\end{figure}

\begin{figure}[h]
	\centering
	\includegraphics[width=0.6\textwidth]{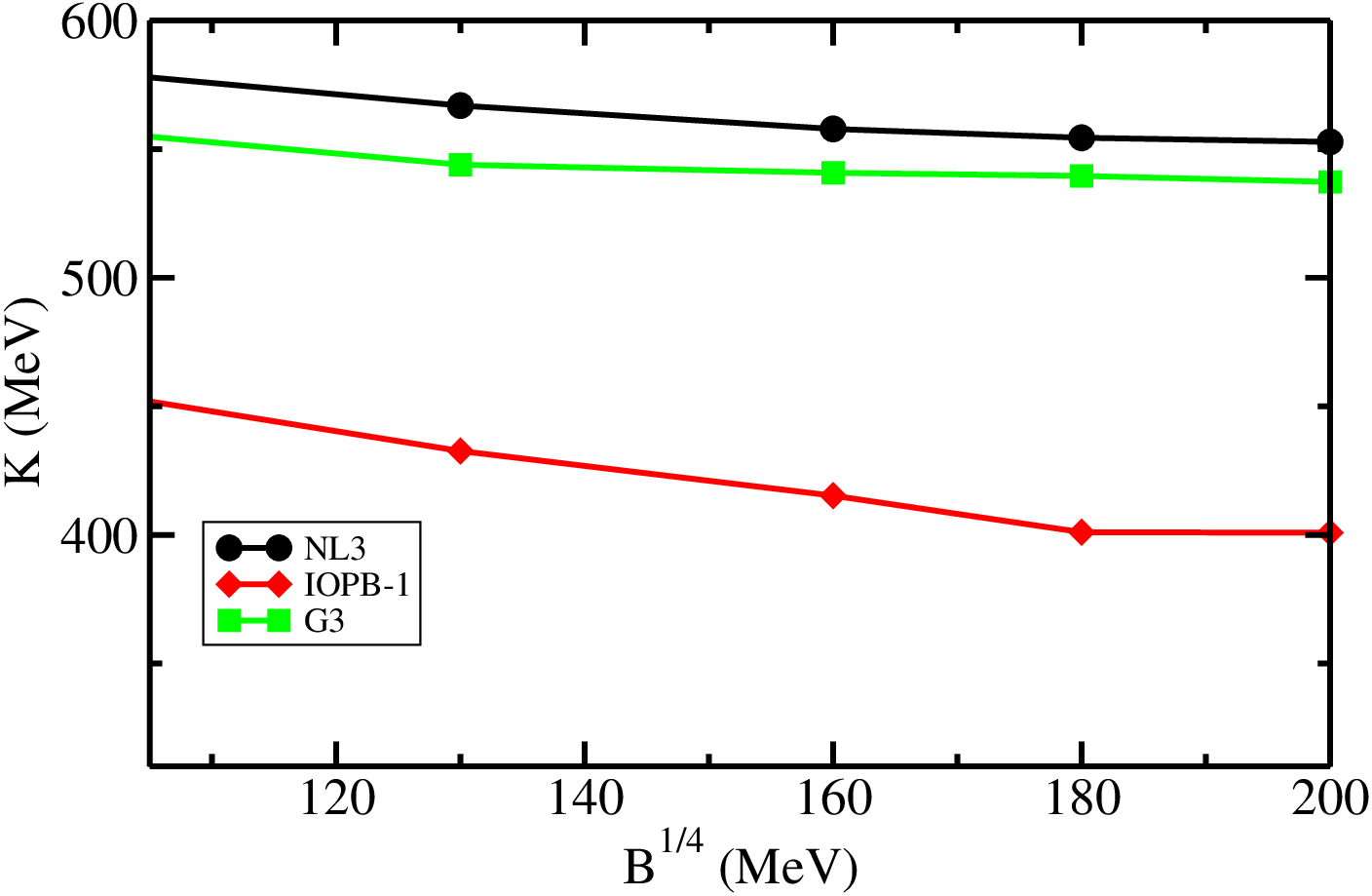}
	\caption{ NM incompressibility for Hybrid EoS as a function of bag constant $B^{1/4}$ for a) NL3, b) IOPB-I and c) G3 parameter sets.}
	\label{fig7}
\end{figure}

 
All the parameters of hybrid EoS like $J$, $L$, $K_{sym}$ and the skewness parameter $Q_{sym}$ at saturation are plotted as a function of Bag constant $B^{1/4}$ for different parameter sets as displayed in figure (\ref{fig6}). The symmetry energy $J$ increases with the bag values for IOPB-I and G3 sets, while for NL3 it decreases initially for $B^{1/4}$ values upto 180 MeV and then increases for 200 MeV, showing a completely different nature than the rest of parameter sets. The slope parameter $L$ and $K_{sym}$ vary almost in the similar fashion for  IOPB-I and G3 sets. NL3 has higher values for both $L$ and $K_{sym}$ whereas the $Q_{sym}$ is large for G3 set. The skewness coefficient of symmetry energy $Q_{sym}$ has very large value for all parameter sets. For G3 set, the values lie in the range 5500-6500 MeV, while for IOPB-I set, it lies within 2000-3000 MeV. For hadronic matter the $Q_{sym}$ is largely uncertain, but its value is predicted from (100-1000) MeV.  The incompressibility coefficient $K$ for all parameter sets is displayed in figure (\ref{fig7}). The values lie in the range 400-600 MeV, which is very large compared to the predicted values from ISGMR \cite{Colo:2013yta,Piekarewicz:2013bea}.The values of symmetry energy and all other quantities are very high compared to the values of the pure hadronic matter. All these quantities cannot be measured directly by experiments and the theoretical models produce a very wide range of such values. So, there is always uncertainty associated with the measurement of such quantities and hence the calculations of such values in this regard is more needed. A good knowledge of the EoS and better understanding on the theoretical models can help in determining these quantities with a great accuracy.

\begin{figure}[h]
	\centering
	\includegraphics[width=0.6\textwidth]{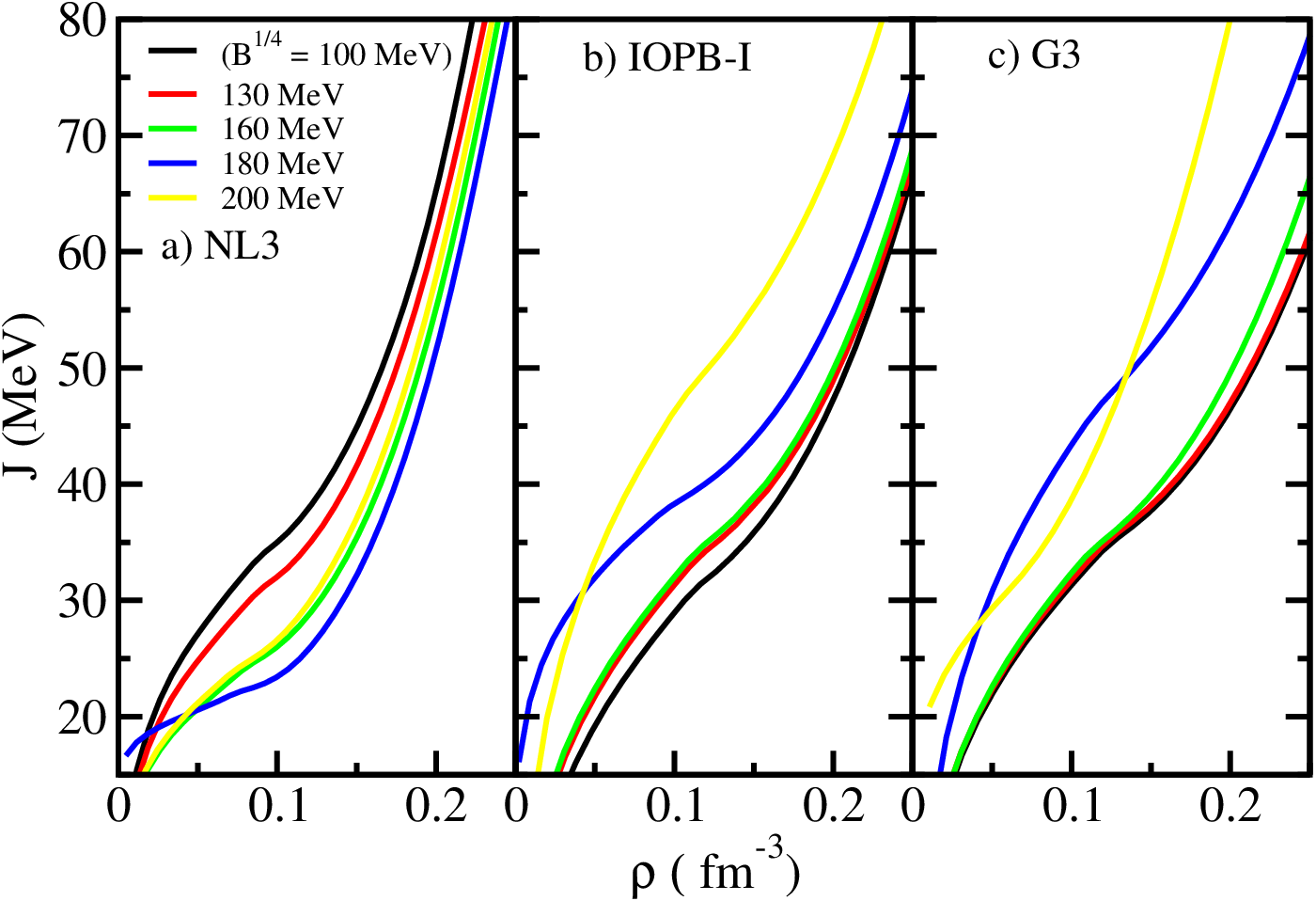}
	\caption{ Symmetry energy $J$ versus density for hybrid EoS with different bag values for a) NL3, b) IOPB-I and c) G3 parameter sets.}
	\label{fig8}
\end{figure}

The variation of symmetry energy for hybrid EoS with density for different HM parameter sets and different bag values are shown in figure (\ref{fig8}).
The symmetry energy for NL3 set increases smoothly with density for all bag constants. However for IOPB-I and G3 sets, the symmetry energy shows a rapid increase for bag constants $B^{1/4}$ =180 and 200 MeV. The G3 set produces softer symmetry energy for low bag values in comparison to the IOPB-I and NL3 sets, while as it produces very stiff value of symmetry energy for bag constants 180 and 200 MeV. The symmetry energy at saturation density $\rho_0$ for NL3 set initially decreases with bag constant upto $B^{1/4}$=180 MeV, thereafter it increases for $B^{1/4}$=200 MeV.  No such variation in the symmetry energy is seen for IOPB-I and G3 sets. The large variation in symmetry energy for 180 MeV and 200 MeV bag values for IOPB-I and G3 sets at higher densities may well contribute to the star matter properties.\\

\begin{figure}[h]
	\centering
	\includegraphics[width=0.6\textwidth]{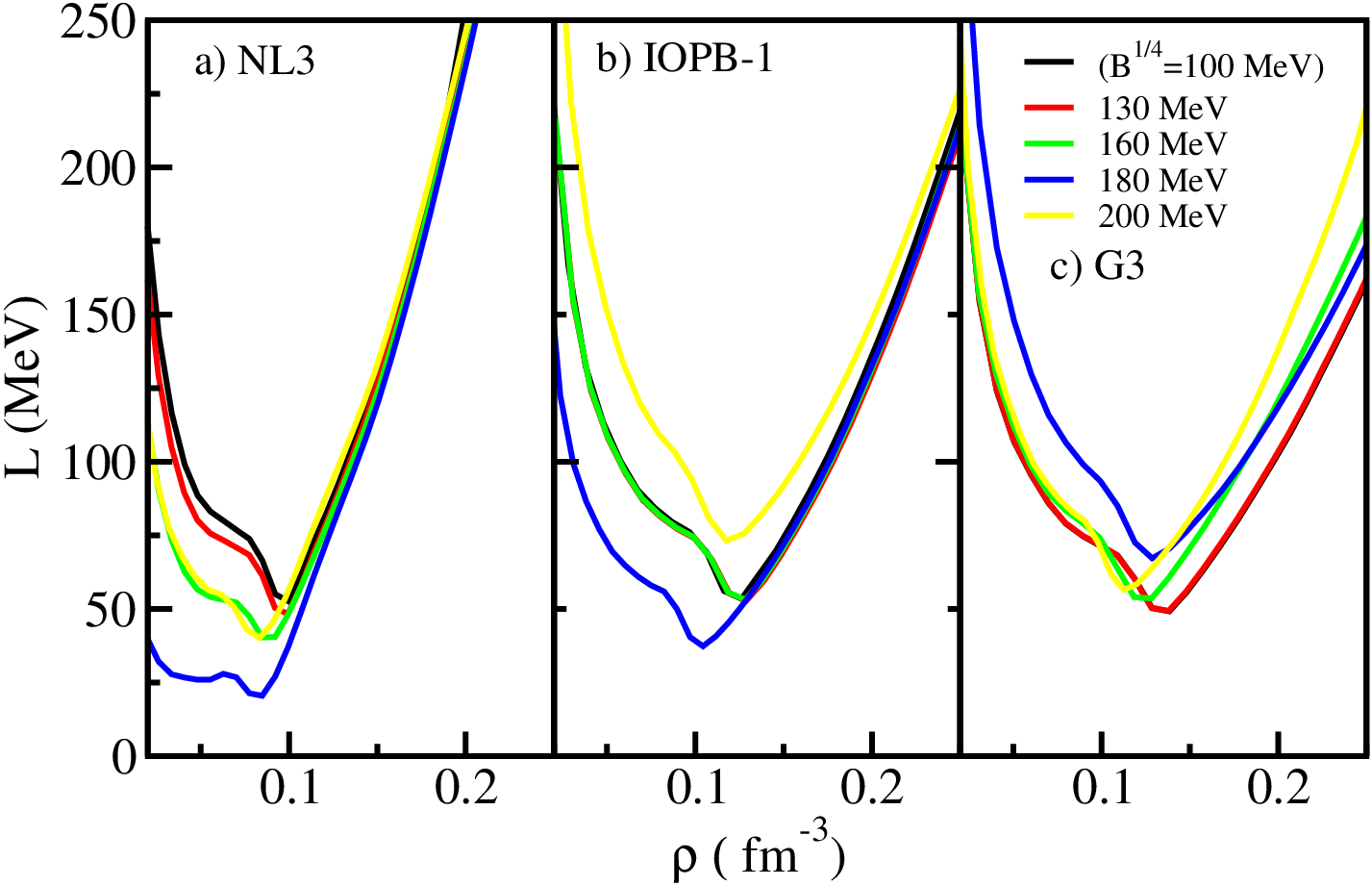}
		\caption{ Slope parameter $L$ as a function of density for hybrid EoS with different bag values for a) NL3, b) IOPB-I and c) G3 parameter sets.}
		\label{fig9}
\end{figure}

The slope parameter $L$ versus $\rho$ is displayed in figure (\ref{fig9}). For NL3 set, $L$ shows similar behavior for all bag constants at densities $\rho$ $>$ $\rho_0$. However, for densities $\rho$ $<$ $\rho_0$, the $L$ value shows more saturation for all bag constants. IOPB-I set follows almost similar pattern. However for G3 set, the $L$ value increases with density at $\rho$ $>$ $\rho_0$. The G3 set produces soft slope parameter $L$ as compared to NL3 and IOPB-I parameterizations.\par 

Now with the EoS's obtained for the hybrid stars, we use the Tolman-Oppenheimer-Volkoff (TOV) equations \cite{PhysRev.55.374,PhysRev.55.364} that are used to evaluate the structure of the star. Assuming the star to be spherical and stationary, we have\\

\begin{equation}\label{tov1}
\frac{dP(r)}{dr}= -G\frac{[\mathcal{E}(r) +P(r)][M(r)+4\pi r^3 P(r)]}{r^2(1-2M(r)/r) }
\end{equation}
and
\begin{equation}\label{tov2}
\frac{dM(r)}{dr}= 4\pi r^2 \mathcal{E}(r)
\end{equation}
Here, $G$ is the gravitational constant and $M(r)$ is the gravitational mass. For a given EoS, the equations (\ref{tov1}) and (\ref{tov2}) are solved for the given boundary conditions $P(0)=P_c$, $M(0)=0$, where $P_c$ is the central pressure. \par 

\begin{figure}
	
	\begin{minipage}{.5\linewidth}
		\centering
		\subfloat[]{\label{main:a}\includegraphics[scale=.27]{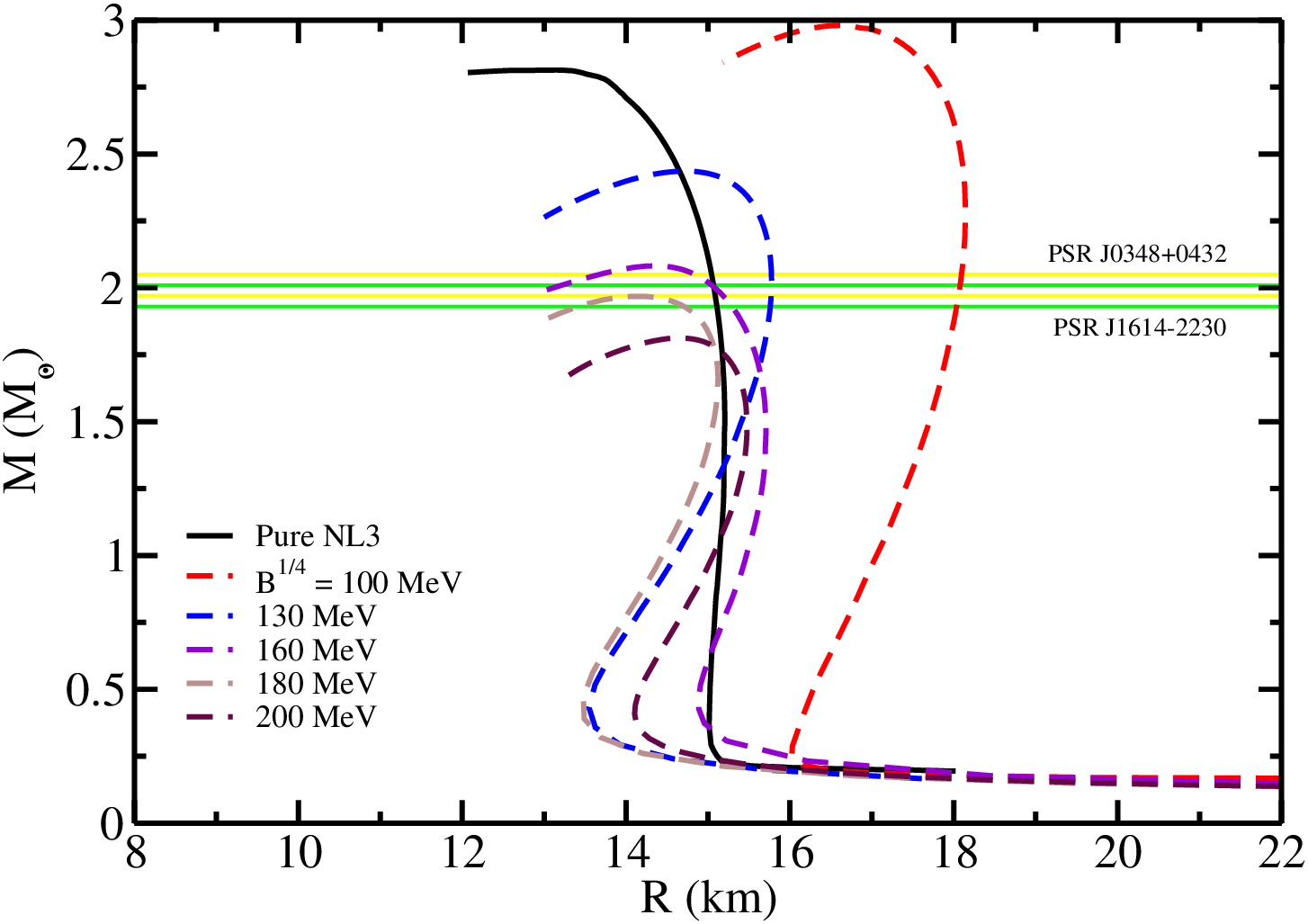}}
	\end{minipage}%
	\begin{minipage}{.5\linewidth}
		\centering
		\subfloat[]{\label{main:b}\includegraphics[scale=.27]{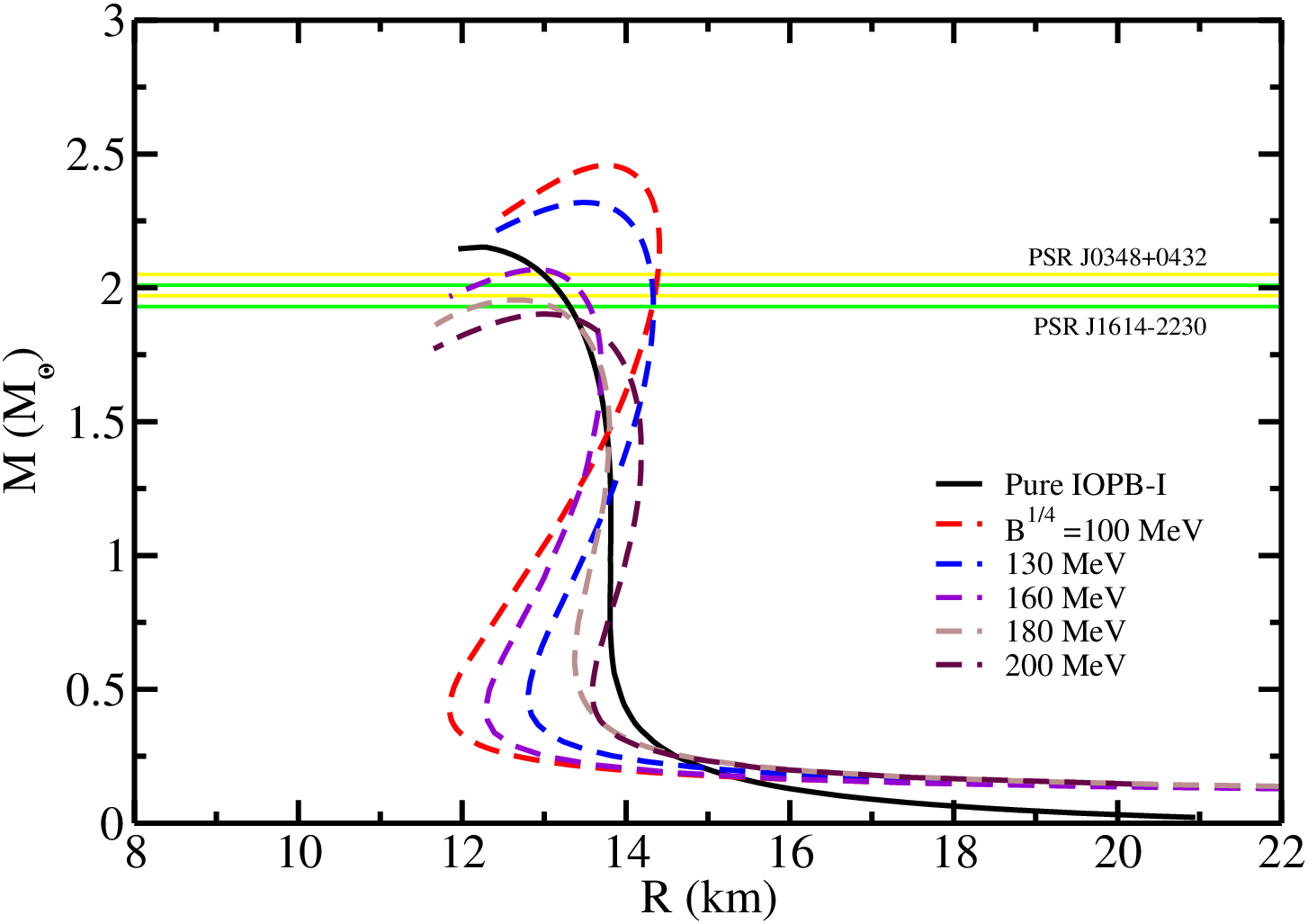}}
	\end{minipage}\par\medskip
	\centering
	\subfloat[]{\label{main:c}\includegraphics[scale=.3]{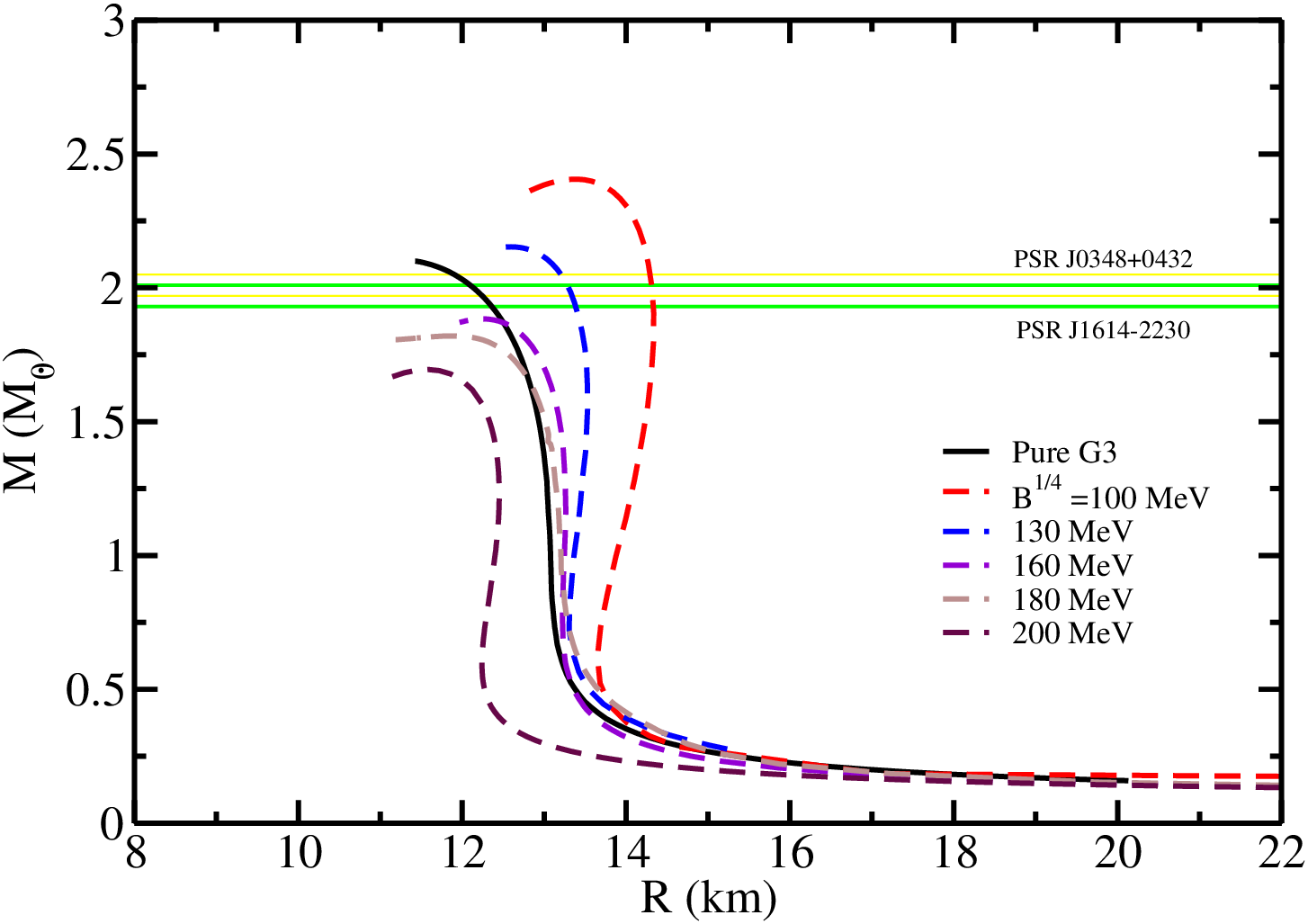}}
	
	\caption{Mass Radius profile of hybrid star for a) NL3, b) IOPB-I and c) G3 parameter sets \cite{rather2020constraining}. The recent observational constraints on the maximum mass and radius \cite{2010Natur.467.1081D,Antoniadis1233232} are also shown. }
	\label{hyb}
\end{figure}

The maximum mass of the neutron star with pure hadron matter for NL3, IOPB-I and G3 parameter sets is 2.81, 2.15 and 2.03 $M_{\odot}$ respectively. With the addition of quarks, the maximum mass of hybrid stars decreases from 2.40 to 1.69 $M_{\odot}$ for G3 set as the bag constant increases from 100 to 200 MeV. Similarly for NL3 and IOPB-I parameter sets, the maximum mass decreases from 2.98 to 1.81 $M_{\odot}$ and 2.46 to 1.90 $M_{\odot}$, respectively. The recent observational constraints on the maximum mass are also shown. The green band represents the precisely measured mass of binary millisecond pulsar PSR J1614-2230 (1.97$\pm$0.04 $M_{\odot}$) \cite{2010Natur.467.1081D}, while the yellow band represents the measured mass of PSR J0348+0432  (2.01$\pm$0.04 $M_{\odot}$) \cite{Antoniadis1233232}. Pure hadron matter with NL3 set produces the maximum mass of a star which is usually ruled out by the recent observational constraints. However, the addition of quarks with proper bag constants reduces the maximum mass well within the limits.\par 
The recently measured gravitational wave observation of a binary neutron star inspiral GW170817 constrains the neutron star maximum mass \cite{PhysRevLett.119.161101}. Combining the results from GW170817 and the quasi-universal relation between rotating and non-rotating neutron stars, the maximum mass of a non-rotating neutron star is found to be in the range 2.01$\pm$0.04$<M (M_{\odot})<$2.16$\pm$0.03 \cite{Rezzolla_2018}. The most recent gravitational wave observation GW190425 constrains the NS maximum mass in the range 1.12 to 2.52 M ($M_{\odot}$) \cite{Abbott_2020}. Comparing our calculated maximum mass results with these measured GW data, we find that except for bag values B$^{1/4}$ =100 and 200 MeV, the maximum mass obtained for hybrid neutron stars satisfies the GW limits.\\  

The variation in all these nuclear matter properties with bag constant will be important in the context of constraining the EoS of nuclear matter. Such variations will directly effect the star matter properties. Furthermore, it may allow us to properly calculate the fraction of quark matter present in neutron stars. Considering color flavor or one gluon exchange in the simple MIT bag model or using other models like NJL \cite{PhysRev.122.345,PhysRev.124.246} for quark matter may further constrain these nuclear matter properties for hybrid EoS.\par 

Since the symmetry energy cannot be measured directly, it is important to identify the observables that correlate  the symmetry energy  and its density dependence to impose constraints on the quantities like slope parameter, symmetry energy curvature etc. The additional information about these quantities can be extracted from the astrophysical observations of high dense matter objects like neutron stars or from a better knowledge of EoS. The nature of EoS is influenced remarkably with these quantities and since these parameters are controlled by the bag constant $B$, then it will be possible to adjust the mass and radius of neutron star by tuning the bag constant $B$.\\
\section{Summary and Conclusion}
We studied the hybrid EoS by mixing hadron matter and quark matter using Gibbs conditions. E-RMF model for hadron matter with recently reported parameter sets and MIT bag model for quark matter with different bag constants is studied. The nuclear matter properties like symmetry energy ($J$), slope parameter ($L$), curvature of symmetry energy ($K_{sym}$), skewness ($Q_{sym}$) and incompressibility ($K$) are calculated for hybrid EoS. The MR relation for all the EoS's is also obtained and it is found that the maximum mass of a star decreases as the bag constant varies from (100-200) MeV. It is found that the values of symmetry energy $J$ and other quantities are very high for a hybrid EoS and they increase with the bag constant except the $J$  and $L$ values (for NL3) and  incompressibility $K$ (for all parameter sets) which decreases with the bag values $B$. The values obtained for symmetry energy and other quantities are very large as compared to their predicted values for hadronic matter. The predicted values of these quantities from various theoretical models also have a large uncertainty. \par 
All these quantities have a huge impact on the neutron star mass-radius relation and other important quantities. The slope parameter has influence on the properties of both finite and infinite nuclear matter. The phase transition properties of hadron-quark matter and the existence of exotic phases like Kaons, Hyperons etc. in neutron stars are also dependent on these quantities. The effect of temperature on symmetry energy and other quantities will allow us to understand the compact objects more deeper. The presence of quarks inside the neutron star core modifies the equation of state and changes the nuclear matter properties. It will be interesting to see how these values for a hybrid star will provide a new insight into the physics of neutron stars and other high dense objects.

\section*{Acknowledgment}
Ishfaq A. Rather is thankful to the Institute of Physics, Bhubaneswar for providing the hospitality during the work.
\section*{References}
\bibliographystyle{iopart-num}
\bibliography{iopart-num}

\providecommand{\newblock}{}
\begin{thebibliography}{10}
\expandafter\ifx\csname url\endcsname\relax
  \def\url#1{{\tt #1}}\fi
\expandafter\ifx\csname urlprefix\endcsname\relax\def\urlprefix{URL }\fi
\providecommand{\eprint}[2][]{\url{#2}}

\bibitem{PhysRevD.30.272}
Witten E 1984 {\em Phys. Rev. D\/} {\bf \textbf{30}}(2) 272

\bibitem{PhysRevD.30.2379}
Farhi E and Jaffe R~L 1984 {\em Phys. Rev. D\/} {\bf \textbf{30}}(11) 2379

\bibitem{PhysRevD.46.1274}
Glendenning N~K 1992 {\em Phys. Rev. D\/} {\bf \textbf{46}}(4) 1274

\bibitem{Danielewicz1592}
Danielewicz P, Lacey R and Lynch W~G 2002 {\em Science\/} {\bf \textbf{298}}
  1592

\bibitem{Lattimer:2015nhk}
Lattimer J~M and Prakash M 2016 {\em Phys. Rep.\/} {\bf \textbf{621}} 127

\bibitem{RevModPhys.88.021001}
Watts A~L~e~a 2016 {\em Rev. Mod. Phys.\/} {\bf \textbf{88}}(2) 021001

\bibitem{RevModPhys.89.015007}
Oertel M, Hempel M, Kl\"ahn T and Typel S 2017 {\em Rev. Mod. Phys.\/} {\bf
  \textbf{89}}(1) 015007

\bibitem{Ozel:2016oaf}
\"Ozel F and Freire P 2016 {\em Ann. Rev. Astrophys.\/} {\bf \textbf{54}} 401

\bibitem{PhysRevC.85.035201}
Dutra M, Louren\ifmmode~\mbox{\c{c}}\else \c{c}\fi{}o O, S\'a~Martins J~S,
  Delfino A, Stone J~R and Stevenson P~D 2012 {\em Phys. Rev. C\/} {\bf
  \textbf{85}}(3) 035201

\bibitem{PhysRevC.90.055203}
Dutra M~e~a 2014 {\em Phys. Rev. C\/} {\bf \textbf{90}}(5) 055203

\bibitem{PhysRevLett.106.252501}
Roca-Maza X, Centelles M, Vi\~nas X and Warda M 2011 {\em Phys. Rev. Lett.\/}
  {\bf \textbf{106}}(25) 252501

\bibitem{PhysRevC.98.065801}
Wu X, Ohnishi A and Shen H 2018 {\em Phys. Rev. C\/} {\bf 98}(6) 065801

\bibitem{THORSSON1994693}
Thorsson V, Prakash M and Lattimer J~M 1994 {\em Nucl. Phys. A\/} {\bf 572} 693
  -- 731

\bibitem{PhysRevC.95.015801}
Kumar B, Biswal S~K and Patra S~K 2017 {\em Phys. Rev. C\/} {\bf 95}(1) 015801

\bibitem{FURNSTAHL1996539}
Furnstahl R, Serot B~D and Tang H~B 1996 {\em Nucl. Phys. A\/} {\bf
  \textbf{598}} 539

\bibitem{PhysRevD.9.3471}
Chodos A, Jaffe R~L, Johnson K, Thorn C~B and Weisskopf V~F 1974 {\em Phys.
  Rev. D\/} {\bf \textbf{9}}(12) 3471

\bibitem{PhysRevD.17.1109}
Freedman B and McLerran L 1978 {\em Phys. Rev. D\/} {\bf \textbf{17}}(4) 1109

\bibitem{Walecka:1974qa}
Walecka J~D 1974 {\em Ann. Phys.\/} {\bf \textbf{83}} 491

\bibitem{Reinhard:1989zi}
Reinhard P~G 1989 {\em Rept. Prog. Phys.\/} {\bf \textbf{52}} 439

\bibitem{Serot_1992}
Serot B~D 1992 {\em Rep. Prog. Phys.\/} {\bf \textbf{55}} 1855

\bibitem{Horowitz:1981xw}
Horowitz C~J and Serot B~D 1981 {\em Nucl. Phys. A\/} {\bf 368} 503--528

\bibitem{BOGUTA1977413}
Boguta J and Bodmer A 1977 {\em Nucl. Phys. A\/} {\bf \textbf{292}} 413

\bibitem{GAMBHIR1990132}
Gambhir Y, Ring P and Thimet A 1990 {\em Ann. Phys.\/} {\bf \textbf{198}} 132

\bibitem{RING1996193}
Ring P 1996 {\em Prog. Part. and Nucl. Phys.\/} {\bf \textbf{37}} 193

\bibitem{Arumugam:2004ys}
Arumugam P, Sharma B~K, Sahu P~K, Patra S~K, Sil T, Centelles M and Vinas X
  2004 {\em Phys. Lett. B\/} {\bf \textbf{601}} 51

\bibitem{PhysRevC.55.540}
Lalazissis G~A, K\"onig J and Ring P 1997 {\em Phys. Rev. C\/} {\bf
  \textbf{55}}(1) 540

\bibitem{Gambhir:1989mp}
Gambhir Y~K, Ring P and Thimet A 1990 {\em Ann. Phys.\/} {\bf \textbf{198}} 132

\bibitem{BODMER1991703}
 1991 {\em Nucl.. Phys. A\/} {\bf \textbf{526}} 703

\bibitem{PhysRevC.68.054318}
Bunta J~K and Gmuca S 2003 {\em Phys. Rev. C\/} {\bf \textbf{68}}(5) 054318

\bibitem{GMUCA1992447}
 1992 {\em Nucl. Phys. A\/} {\bf \textbf{547}} 447

\bibitem{SUGAHARA1994557}
Sugahara Y and Toki H 1994 {\em Nucl. Phys. A\/} {\bf \textbf{579}} 557

\bibitem{PhysRevLett.95.122501}
Todd-Rutel B~G and Piekarewicz J 2005 {\em Phys. Rev. Lett.\/} {\bf
  \textbf{95}}(12) 122501

\bibitem{PhysRevC.82.055803}
Fattoyev F~J, Horowitz C~J, Piekarewicz J and Shen G 2010 {\em Phys. Rev. C\/}
  {\bf \textbf{82}}(5) 055803

\bibitem{KUBIS1997191}
Kubis S and Kutschera M 1997 {\em Phys. Lett. B\/} {\bf \textbf{399}} 191

\bibitem{PhysRevC.89.044001}
Singh S~K, Biswal S~K, Bhuyan M and Patra S~K 2014 {\em Phys. Rev. C\/} {\bf
  \textbf{89}}(4) 044001

\bibitem{Biswal:2016zcg}
Biswal S~K, Kumar B and Patra S~K 2016 {\em Int. J. Mod. Phys. E\/} {\bf 25}
  1650090

\bibitem{Kumara:2017bti}
Kumar B, Singh S~K, Agrawal B~K and Patra S~K 2017 {\em Nucl. Phys. A\/} {\bf
  966} 197--207

\bibitem{PhysRevC.97.045806}
Kumar B, Patra S~K and Agrawal B~K 2018 {\em Phys. Rev. C\/} {\bf
  \textbf{97}}(4) 045806

\bibitem{PhysRevC.63.024314}
Del~Estal M, Centelles M, Vi\~nas X and Patra S~K 2001 {\em Phys. Rev. C\/}
  {\bf \textbf{63}}(2) 024314

\bibitem{kapusta_gale_2006}
Kapusta J~I 1989 {\em Finite-Temperature Field Theory\/} 2nd ed (Cambridge
  University Press)

\bibitem{Baym:2017whm}
Baym G, Hatsuda T, Kojo T, Powell P~D, Song Y and Takatsuka T 2018 {\em Rept.
  Prog. Phys.\/} {\bf 81} 056902

\bibitem{STEINER2000239}
Steiner A, Prakash M and Lattimer J 2000 {\em Phys. Lett. B\/} {\bf 486} 239 --
  248

\bibitem{BUBALLA2005205}
Buballa M 2005 {\em Phys. Rep.\/} {\bf 407} 205 -- 376

\bibitem{NOVIKOV1981301}
Novikov V, Shifman M, Vainshtein A and Zakharov V 1981 {\em Nucl. Phys. B\/}
  {\bf 191} 301 -- 369

\bibitem{PhysRevD.22.1198}
Haxton W~C and Heller L 1980 {\em Phys. Rev. D\/} {\bf 22}(5) 1198--1208

\bibitem{PhysRevD.12.2060}
DeGrand T, Jaffe R~L, Johnson K and Kiskis J 1975 {\em Phys. Rev. D\/} {\bf
  12}(7) 2060--2076

\bibitem{rather2020constraining}
Rather I~A, Usmani A~A, Imran M and Patra S~K 2020 (\textit{Preprint}
  \eprint{2002.00616})

\bibitem{PhysRevC.60.025801}
Schertler K, Leupold S and Schaffner-Bielich J 1999 {\em Phys. Rev. C\/} {\bf
  \textbf{60}}(2) 025801

\bibitem{PhysRevC.75.035808}
Sharma B~K, Panda P~K and Patra S~K 2007 {\em Phys. Rev. C\/} {\bf
  \textbf{75}}(3) 035808

\bibitem{PhysRevC.66.025802}
Burgio G~F, Baldo M, Sahu P~K and Schulze H~J 2002 {\em Phys. Rev. C\/} {\bf
  \textbf{66}}(2) 025802

\bibitem{PhysRevC.89.015806}
Orsaria M, Rodrigues H, Weber F and Contrera G~A 2014 {\em Phys. Rev. C\/} {\bf
  \textbf{89}}(1) 015806

\bibitem{PhysRevD.88.063001}
Logoteta D and Bombaci I 2013 {\em Phys. Rev. D\/} {\bf \textbf{88}}(6) 063001

\bibitem{GLENDENNING2001393}
Glendenning N~K 2001 {\em Phys. Rept.\/} {\bf 342} 393 -- 447

\bibitem{FARINE1978317}
Farine M, Pearson J and Rouben B 1978 {\em Nucl. Phys. A\/} {\bf \textbf{304}}
  317

\bibitem{PEARSON19911}
Pearson J, Aboussir Y, Dutta A, Nayak R, Farine M and Tondeur F 1991 {\em Nucl.
  Phys. A\/} {\bf \textbf{528}} 1

\bibitem{PhysRevC.44.1892}
Bombaci I and Lombardo U 1991 {\em Phys. Rev. C\/} {\bf \textbf{44}}(5) 1892

\bibitem{PhysRevC.84.054309}
Roca-Maza X, Vi\~nas X, Centelles M, Ring P and Schuck P 2011 {\em Phys. Rev.
  C\/} {\bf \textbf{84}}(5) 054309

\bibitem{Singh_2013}
Singh S~K, Bhuyan M, Panda P~K and Patra S~K 2013 {\em J. of Phys. G: Nucl. and
  Part. Phys.\/} {\bf \textbf{40}} 085104

\bibitem{PhysRevC.86.015803}
Tsang M~B~e~a 2012 {\em Phys. Rev. C\/} {\bf \textbf{86}}(1) 015803

\bibitem{PhysRevC.82.054607}
Xu C, Li B~A and Chen L~W 2010 {\em Phys. Rev. C\/} {\bf \textbf{82}}(5) 054607

\bibitem{Newton_2012}
Newton W~G, Gearheart M and Li B~A 2012 {\em The Astrophys. J.\/} {\bf
  \textbf{204}} 9

\bibitem{PhysRevLett.108.081102}
Steiner A~W and Gandolfi S 2012 {\em Phys. Rev. Lett.\/} {\bf \textbf{108}}(8)
  081102

\bibitem{PhysRevC.86.025804}
Fattoyev F~J, Newton W~G, Xu J and Li B~A 2012 {\em Phys. Rev. C\/} {\bf
  \textbf{86}}(2) 025804

\bibitem{PhysRevLett.102.122502}
Centelles M, Roca-Maza X, Vi\~nas X and Warda M 2009 {\em Phys. Rev. Lett.\/}
  {\bf \textbf{102}}(12) 122502

\bibitem{LI2013276}
Li B~A and Han X 2013 {\em Phys. Lett. B\/} {\bf \textbf{727}} 276

\bibitem{Lattimer_2001}
Lattimer J~M and Prakash M 2001 {\em The Astrophys. J.\/} {\bf \textbf{550}}
  426

\bibitem{FURNSTAHL200285}
Furnstahl R 2002 {\em Nucl. Phys. A\/} {\bf \textbf{706}} 85

\bibitem{PhysRevLett.86.5647}
Horowitz C~J and Piekarewicz J 2001 {\em Phys. Rev. Lett.\/} {\bf
  \textbf{86}}(25) 5647

\bibitem{PhysRevC.97.025806}
Margueron J, Hoffmann~Casali R and Gulminelli F 2018 {\em Phys. Rev. C\/} {\bf
  \textbf{97}}(2) 025806

\bibitem{CHEN2015284}
Chen W~C and Piekarewicz J 2015 {\em Phys. Lett. B\/} {\bf \textbf{748}} 284

\bibitem{KUMAR2017197}
Kumar B, Singh S, Agrawal B and Patra S 2017 {\em Nucl. Phys. A\/} {\bf
  \textbf{966}} 197

\bibitem{Colo:2013yta}
Colo G, Garg U and Sagawa H 2014 {\em Eur. Phys. J. A\/} {\bf \textbf{50}} 26

\bibitem{Piekarewicz:2013bea}
Piekarewicz J 2014 {\em Eur. Phys. J. A\/} {\bf \textbf{50}} 25

\bibitem{Ghosh1995}
Ghosh S~K, Phatak S~C and Sahu P~K 1995 {\em Phys. A Had. and Nucl.\/} {\bf
  352}(7) 2060--2076

\bibitem{PhysRev.55.374}
Oppenheimer J~R and Volkoff G~M 1939 {\em Phys. Rev.\/} {\bf 55}(4) 374--381

\bibitem{PhysRev.55.364}
Tolman R~C 1939 {\em Phys. Rev.\/} {\bf 55}(4) 364--373

\bibitem{2010Natur.467.1081D}
Demorest P~B, Pennucci T, Ransom S~M, Roberts M~S~E and Hessels J~W~T 2010 {\em
  Nature\/} {\bf 467} 1081

\bibitem{Antoniadis1233232}
Antoniadis J~e~a 2013 {\em Science\/} {\bf 340}

\bibitem{PhysRevLett.119.161101}
Abbott B~P and Abbott R (LIGO Scientific Collaboration and Virgo Collaboration)
  2017 {\em Phys. Rev. Lett.\/} {\bf 119}(16) 161101
  \urlprefix\url{https://link.aps.org/doi/10.1103/PhysRevLett.119.161101}

\bibitem{Rezzolla_2018}
Rezzolla L, Most E~R and Weih L~R 2018 {\em The Astrophysical Journal\/} {\bf
  852} L25

\bibitem{Abbott_2020}
Abbott B~P and Abbott R 2020 {\em The Astrophysical Journal\/} {\bf 892} L3

\bibitem{PhysRev.122.345}
Nambu Y and Jona-Lasinio G 1961 {\em Phys. Rev.\/} {\bf 122}(1) 345--358

\bibitem{PhysRev.124.246}
Nambu Y and Jona-Lasinio G 1961 {\em Phys. Rev.\/} {\bf 124}(1) 246--254

\end{thebibliography}
\end{document}